\documentclass[10pt,twocolumn,twoside] {IEEEtran}
\usepackage{graphicx,amsfonts}
\usepackage{subfig,color}
\usepackage{cite}
\ifCLASSINFOpdf
\else
\fi
\usepackage[cmex10]{amsmath}
\usepackage{array,multirow}
\usepackage{mdwmath}
\usepackage{mdwtab}
\DeclareMathOperator*{\argmin}{arg\,min}

\begin{document}

\title{Supervised and Unsupervised Speech Enhancement\\ Using Nonnegative Matrix Factorization}

\author{Nasser~Mohammadiha,
        Paris~Smaragdis,
        Arne~Leijon
\thanks{
This paper is published at at IEEE Transactions on Audio, Speech, and Language Processing, Nov. 2013: http://ieeexplore.ieee.org/document/6544586/}}

\markboth{Published at IEEE Transactions on Audio, Speech, and Language Processing, Nov. 2013}%
{Mohammadiha \MakeLowercase{\textit{et al.}}: Speech Enhancement Using NMF}

\maketitle

\begin{abstract}
Reducing the interference noise in a monaural noisy speech signal has been a challenging task for many years. Compared to traditional
unsupervised speech enhancement methods, e.g., Wiener filtering, supervised approaches, such as algorithms based on hidden Markov models
(HMM), lead to higher-quality enhanced speech signals. However, the main practical difficulty of these approaches is that for each
noise type a model is required to be trained a priori. In this paper, we investigate a new class of supervised speech denoising algorithms
using nonnegative matrix factorization (NMF). We propose a novel speech enhancement method that is based on a Bayesian formulation of NMF (BNMF). To circumvent the mismatch problem between the training and testing stages, we propose two solutions. First, we use an HMM in combination with
BNMF (BNMF-HMM) to derive a minimum mean square error (MMSE) estimator for the speech signal with no information about the underlying noise
type. Second, we suggest a scheme to learn the required noise BNMF model online, which is then used to develop an unsupervised
speech enhancement system. Extensive experiments are carried out to investigate the performance of the proposed methods under different conditions.
Moreover, we compare the performance of the developed algorithms with state-of-the-art speech enhancement schemes using various objective
measures. Our simulations show that the proposed BNMF-based methods outperform the competing algorithms substantially.
\end{abstract}


\IEEEpeerreviewmaketitle

\section{Introduction\label{sec:Introduction}}
Estimating the clean speech signal in a single-channel recording of a noisy speech signal has been a research topic for a long time and is of interest
for various applications including hearing aids, speech/speaker recognition, and speech communication over telephone and internet.
A major outcome of these techniques is the improved quality and reduced listening effort in the presence of an interfering noise signal.

In general, speech enhancement methods can be categorized into two broad classes: unsupervised and supervised. Unsupervised
methods include a wide range of approaches such as spectral subtraction \cite{Boll1979}, Wiener and Kalman filtering, e.g., \cite{Lim1979,Volodya2006}, short-time spectral amplitude (STSA) estimators \cite{Ephraim1984},
estimators based on super-Gaussian prior distributions for speech DFT coefficients \cite{Martin2005,Cohen2006,Erkelens2007,Chen2007}, and
schemes based on periodic models of the speech signal \cite{Jensen2012}. In these methods, a statistical model is assumed for the speech and
noise signals, and the clean speech is estimated from the noisy observation without any prior information on the noise type or speaker
identity. However, the main difficulty of most of these methods is estimation of the noise power spectral density (PSD)  \cite{Martin2001,Cohen2003,Hendriks2010}, which is a challenging task if the background noise is non-stationary.

For the supervised methods, a model is considered for both the speech and noise signals and the model parameters are
estimated using the training samples of that signal. Then, an interaction model is defined by combining speech and noise models and the noise
reduction task is carried out. Some examples of this class of algorithms include the codebook-based approaches, e.g., \cite{Sreenivas1996,Srinivasan2006} and hidden Markov model (HMM) based methods \cite{Ephraim1992,Sameti1998,Zhao2007,Mohammadiha2013a,Veisi2013}. One advantage of these methods is that there is no need to estimate
the noise PSD using a separate algorithm.

The supervised approaches have been shown to produce better quality enhanced speech signals compared to the unsupervised methods \cite{Sameti1998,Srinivasan2006},
which can be expected as more prior information is fed to the system in these cases and the considered models are trained for each specific
type of signals. The required prior information on noise type (and speaker identity in some cases) can be given by the user, or can be
obtained using a built-in classification scheme \cite{Sameti1998,Srinivasan2006}, or can be provided by a separate acoustic environment classification
algorithm \cite{El-Maleh1999}. The primary goal of this work is to propose supervised and unsupervised speech enhancement algorithms
based on nonnegative matrix factorization (NMF) \cite{Lee2000,Cichocki2009}.

NMF is a technique to project a nonnegative matrix $\mathbf{y}$ onto a space spanned by a linear combination of a set of basis vectors, i.e., $\mathbf{y}\approx\mathbf{b}\mathbf{v}$, where both $\mathbf{b}$ and $\mathbf{v}$ are nonnegative matrices.
In speech processing, $\mathbf{y}$ is usually the spectrogram of the speech signal with spectral vectors stored by column,
$\mathbf{b}$ is the basis matrix or dictionary, and $\mathbf{v}$ is referred to as the NMF coefficient or activation matrix. NMF has
been widely used as a source separation technique applied to monaural mixtures, e.g., \cite{Smaragdis2007,Virtanen2007,Fevotte2009}.
More recently, NMF has also been used to estimate the clean speech from a noisy observation \cite{Mohammadiha2013c,Wilson2008b,Schmidt2008a,Mohammadiha2011a,Mysore2011,Mohammadiha2012a}.

When applied to speech source separation, a good separation can be expected only when speaker-dependent basis are learned. In contrast,
for noise reduction, even if a general speaker-independent basis matrix of speech is learned, a good enhancement can be achieved \cite{Mohammadiha2011a,Mohammadiha2012a}. Nevertheless, there might be some scenarios (such as speech degraded with multitalker babble noise) for which the basis matrices of speech and noise are quite similar. In these cases, although the traditional NMF-based approaches can be used to get state-of-the-art
performance, other constraints can be imposed into NMF to obtain a better noise reduction. For instance, assuming that the babble waveform is
obtained as a sum of different speech signals, a nonnegative hidden Markov model is proposed in \cite{Mohammadiha2013c} to model the
babble noise in which the babble basis is identical to the speech basis.  Another fundamental issue in basic NMF is that it ignores the important
temporal dependencies of the audio signals. Different approaches have been proposed in the literature to employ temporal dynamics in NMF, e.g., \cite{Smaragdis2007,Virtanen2007,Wilson2008b,Fevotte2009,Mysore2011,Mohammadiha2012a}.

In this paper, we first propose a new supervised NMF-based speech enhancement system. In the proposed method, the temporal dependencies of speech and noise signals are used to construct informative prior distributions that are applied in a Bayesian framework to perform NMF (BNMF). We then develop an HMM structure with output density functions given by BNMF to simultaneously classify the environmental noise and enhance the noisy signal. Therefore, the noise type doesn't need to be specified a priori. Here, the classification is done using the noisy input and is not restricted to be applied at only the speech pauses as it is in \cite{Sameti1998}, and it doesn't require any additional noise PSD tracking algorithm, as it is required in \cite{Srinivasan2006}.

Moreover, we propose an unsupervised NMF-based approach in which the noise basis matrix is learned online from the noisy mixture. Although online dictionary learning from clean data has been addressed in some prior works, e.g., \cite{Mairal2010,Lefevre2011}, our causal method learns the noise basis matrix from the noisy mixture. The main contributions of this work can be summarized as:
\begin{enumerate}
\item We present a review of state-of-the-art NMF-based noise reduction approaches.
\item We propose a speech enhancement method based on BNMF that inherently captures the temporal dependencies in the form of hierarchical prior distributions. Some preliminary results of this approach has been presented in \cite{Mohammadiha2012a}. Here, we further develop the method and evaluate its performance comprehensively. In particular, we present an approach to construct SNR-dependent prior distributions.
\item An environmental noise classification technique is suggested and is combined with the above BNMF approach (BNMF-HMM) to develop an unsupervised speech enhancement system.
\item A causal online dictionary learning scheme is proposed that learns the noise basis matrix from the noisy observation. Our simulations show that the final unsupervised noise reduction system outperforms state-of-the-art approaches significantly.
\end{enumerate}
The rest of the paper is organized as follows: The review of the NMF-based speech enhancement algorithms is presented in Section \ref{sec:Review-of-State-of-the-art}. In Section \ref{sec:Speech-Enhancement-Using}, we describe our main contributions, namely the BNMF-based noise reduction, BNMF-HMM structure, and online noise dictionary learning. Section \ref{sec:Experiments-and-Results} presents our experiments and results with supervised and unsupervised noise reduction systems. Finally, Section \ref{sec:Conclusions} concludes the study.
\section{Review of State-of-the-art NMF-Based Speech Enhancement\label{sec:Review-of-State-of-the-art}}
In this section, we first explain a basic NMF approach, and then we review NMF-based speech enhancement. Let us represent
the random variables associated with the magnitude of the discrete Fourier transform (DFT) coefficients of the speech, noise, and noisy signals as $\mathbf{S}=[S_{kt}]$, $\mathbf{N}=[N_{kt}]$
and $\mathbf{Y}=[Y_{kt}]$, respectively, where $k$ and $t$ denote the frequency and time indices, respectively. The actual realizations
are shown in small letters, e.g., $\mathbf{y}=[y_{kt}]$. Table \ref{tab:notations} summarizes some of the notations that
are frequently used in the paper.
\begin{table}
\caption{\label{tab:notations}The table summarizes some of the notations that are consistently used in the paper.}

\begin{tabular}{ll}
$k$ & frequency index\tabularnewline
$t$ & time index\tabularnewline
$X$ & a scalar random variable \tabularnewline
$\mathbf{Y}=[Y_{kt}]$ & a matrix of random variabels\tabularnewline
$\mathbf{Y}_{t}$ & $t$-th column of $\mathbf{Y}$\tabularnewline
$\mathbf{y}=[y_{kt}]$ & a matrix of observed magnitude spectrogram\tabularnewline
$\mathbf{y}_{t}$ & $t$-th column of $\mathbf{y}$\tabularnewline
$\mathbf{{b}}^{\left(s\right)}$ & speech parameters ($\mathbf{{b}}^{\left(s\right)}$ is the speech basis matrix)\tabularnewline
${\mathbf{b}}^{\left(n\right)}$ & noise parameters (${\mathbf{b}}^{\left(n\right)}$ is the noise basis matrix)\tabularnewline
$\mathbf{b}=\left[\mathbf{{b}}^{\left(s\right)}\;{\mathbf{b}}^{\left(n\right)}\right]$ & mixture parameters ($\mathbf{b}$ is the mixture basis matrix)\tabularnewline
\end{tabular}
\end{table}
\indent To obtain a nonnegative decomposition of a given matrix $\mathbf{x}$, a cost
function is usually defined and is minimized. Let us denote the basis matrix and NMF coefficient matrix by $\mathbf{b}$ and $\mathbf{v}$,
respectively. Nonnegative factorization is achieved by solving the following optimization problem:
\begin{eqnarray}
\left(\mathbf{b},\mathbf{v}\right) & = & \argmin_{\mathbf{b},\mathbf{v}}\;\: D(\mathbf{y}\|\mathbf{b}\mathbf{v})+\mu h\left(\mathbf{b},\mathbf{v}\right),
\label{eq:NMF_costFunction}
\end{eqnarray}
where $D(\mathbf{y}\|\hat{\mathbf{y}})$ is a cost function, $h(\cdot)$ is an optional regularization term, and $\mu$ is the regularization
weight. The minimization in (\ref{eq:NMF_costFunction}) is performed under the nonnegativity constraint of $\mathbf{b}$ and $\mathbf{v}$.
The common choices for the cost function include Euclidean distance \cite{Lee2000}, generalized Kullback-Leibler divergence \cite{Lee2000,Cemgil2009},
Itakura-Saito divergence \cite{Fevotte2009}, and the negative likelihood of data in the probabilistic NMFs \cite{Smaragdis2006}.
Depending on the application, the sparsity of the activations $\mathbf{v}$ and the temporal dependencies of input data $\mathbf{x}$ are two popular
motivations to design the regularization function, e.g., \cite{Hoyer2004,Virtanen2007,Wilson2008b,Mohammadiha2011b}. Since (\ref{eq:NMF_costFunction}) is not a convex problem, iterative gradient descent or expectation-maximization (EM) algorithms are usually followed to obtain a locally optimal solution for the problem \cite{Lee2000,Smaragdis2006,Fevotte2009}.

Let us consider a supervised denoising approach where the basis matrix of speech ${\mathbf{b}}^{(s)}$ and the basis matrix of noise $\mathbf{{b}}^{(n)}$
are learned using the appropriate training data in advance. The common assumption used to model the noisy speech signal is the
additivity of speech and noise spectrograms, i.e., $\mathbf{\mathbf{y}}=\mathbf{s}+\mathbf{n}$. Although in the real world problems this assumption is not justified completely, the developed algorithms have been shown to produce satisfactory
results, e.g., \cite{Virtanen2007}. The basis matrix of the noisy signal is obtained by concatenating the speech and noise basis matrices
as $\mathbf{b}\hspace{-1mm}=\hspace{-1mm}[\mathbf{{b}}^{(s)}\:\mathbf{{b}}^{(n)}]$. Given the magnitude of DFT coefficients of the noisy speech at time $t$, $\mathbf{y}_{t}$,
the problem in \eqref{eq:NMF_costFunction} is now solved---with  $\mathbf{b}$ held fixed---to obtain the noisy NMF coefficients $\mathbf{v}_{t}$. The NMF decomposition takes the form $\mathbf{y}_{t}\approx\mathbf{b}\mathbf{v}_{t}=[\mathbf{{b}}^{(s)}\:\mathbf{{b}}^{(n)}][({\mathbf{v}}^{(s)}_{t})^{\top}\:\:({\mathbf{v}}^{(n)}_{t})^{\top}]^{\top}$,
where $\top$ denotes transposition. Finally, an estimate of the clean speech DFT magnitudes is obtained by a Wiener-type filtering as:
\begin{equation}
\hat{\mathbf{s}}_{t}=\frac{\mathbf{{b}^{\left(s\right)}}{\mathbf{v}}^{\left(s\right)}_{t}}{\mathbf{{b}}^{\left(s\right)}{\mathbf{v}}^{\left(s\right)}_{t}+\mathbf{{b}}^{\left(n\right)}{\mathbf{v}}^{\left(n\right)}_{t}}\odot\mathbf{y}_{t},
\label{eq:Wiener_filtering}
\end{equation}
where the division is performed element-wise, and $\odot$ denotes an element-wise multiplication. The clean speech waveform is estimated using
the noisy phase and inverse DFT. One advantage of the NMF-based approaches over the HMM-based \cite{Sameti1998,Zhao2007} or codebook-driven
\cite{Srinivasan2006} approaches is that NMF automatically captures the long-term levels of the signals, and no additional gain modeling
is necessary.

Schmidt \textit{et al.} \cite{Schmidt2008a} presented an NMF-based unsupervised batch algorithm for noise reduction. In this approach, it is
assumed that the entire noisy signal is observed, and then the noise basis vectors are learned during the speech pauses. In the intervals of speech
activity, the noise basis matrix is kept fixed and the rest of the parameters (including speech basis and speech and noise NMF coefficients) are
learned by minimizing the Euclidean distance with an additional regularization term to impose sparsity on the NMF coefficients. The enhanced signal is then obtained
similarly to (\ref{eq:Wiener_filtering}). The reported results show that this method outperforms a spectral subtraction algorithm,
especially for highly non-stationary noises. However, the NMF approach is sensitive to the performance of the voice activity detector (VAD). Moreover,
the proposed algorithm in \cite{Schmidt2008a} is applicable only in the batch mode, which is usually not practical in the real world.

In \cite{Wilson2008b}, a supervised NMF-based denoising scheme is proposed in which a heuristic regularization term is added to the cost function. By doing so, the factorization is enforced to follow the pre-obtained statistics. In this method, the basis matrices of speech and noise are learned from training data offline. Also, as part
of the training, the mean and covariance of the log of the NMF coefficients are computed. Using these statistics, the negative likelihood of a
Gaussian distribution (with the calculated mean and covariance) is used to regularize the cost function during the enhancement. The clean
speech signal is then estimated as $\hat{\mathbf{s}}_{t}=\mathbf{{b}}^{(s)}{\mathbf{v}}^{(s)}_{t}$. Although it is not explicitly mentioned in \cite{Wilson2008b}, to make regularization meaningful the statistics of the speech and noise
NMF coefficients have to be adjusted according to the long-term levels of speech and noise signals.

In \cite{Mohammadiha2011a}, authors propose a linear minimum mean square error (MMSE) estimator for NMF-based speech enhancement. In
this work, NMF is applied to $\mathbf{y}_{t}^{p}$ (i.e., $\mathbf{y}_{t}^{p}=\mathbf{b}\mathbf{v}_{t}$, where $p=1$ corresponds to using magnitude of DFT coefficients and $p=2$ corresponds to using magnitude-squared DFT coefficients) in a frame
by frame routine. Then, a gain variable $\mathbf{g}_{t}$ is estimated to filter the noisy signal as: $\hat{\mathbf{s}}_{t}=(\mathbf{g}_{t}\odot\mathbf{y}_{t}^{p})^{1/p}$. Assuming that the basis matrices of speech and noise are obtained
during the training stage, and that the NMF coefficients $\mathbf{V}_{t}$ are random variables, $\mathbf{g}_{t}$ is derived
such that the mean square error between $\mathbf{S}_{t}^{p}$ and $\widehat{\mathbf{S}_{t}^{p}}$ is minimized. The optimal gain is shown to be:
\begin{equation}
\mathbf{g}_{t}=\frac{\boldsymbol{\xi}_{t}+c^{2}\sqrt{\boldsymbol{\xi}_{t}}}{\boldsymbol{\xi}_{t}+1+2c^{2}\sqrt{\boldsymbol{\xi}_{t}}},
\label{eq:LMMSE_gain}
\end{equation}
where $c$ is a constant that depends on $p$ \cite{Mohammadiha2011a} and $\boldsymbol{\xi}_{t}$ is called the smoothed speech to noise ratio that is estimated using
a decision-directed approach. For a theoretical comparison of (\ref{eq:LMMSE_gain}) to a usual Wiener filter see \cite{Mohammadiha2011a}. The conducted simulations show
that the results using $p=1$ are superior to those using $p=2$ (which is in line with previously reported observations, e.g., \cite{Virtanen2007})
and that both of them are better than the results of a state-of-the-art Wiener filter.

A semi-supervised approach is proposed in \cite{Mysore2011} to denoise a noisy signal using NMF. In this method, a nonnegative hidden Markov
model (NHMM) is used to model the speech magnitude spectrogram. Here, the HMM state-dependent output density functions are assumed to be
a mixture of multinomial distributions, and thus, the model is closely related to probabilistic latent component analysis (PLCA) \cite{Smaragdis2006}. An NHMM is described by a set of basis matrices and a Markovian transition matrix that captures the temporal dynamics of the underlying data. To describe a mixture signal, the corresponding NHMMs are then used to construct a factorial HMM. When applied for noise reduction, first a speaker-dependent NHMM is trained on a speech signal. Then, assuming that the whole noisy signal is available (batch mode), the
EM algorithm is run to simultaneously estimate a single-state NHMM for noise and also to estimate the NMF coefficients of the speech and noise signals.
The proposed algorithm doesn't use a VAD to update the noise dictionary, as was done in \cite{Schmidt2008a}. But the algorithm requires the
entire spectrogram of the noisy signal, which makes it difficult for practical applications. Moreover, the employed speech model is speaker-dependent,
and requires a separate speaker identification algorithm in practice. Finally, similar to the other approaches based on the factorial models, the method in \cite{Mysore2011} suffers
from high computational complexity.

A linear nonnegative dynamical system is presented in \cite{Mohammadiha2013b} to model temporal dependencies in NMF. The proposed causal filtering
and fixed-lag smoothing algorithms use Kalman-like prediction in NMF and PLCA. Compared to the ad-hoc methods that
use temporal correlations to design regularity functions, e.g., \cite{Wilson2008b,Mohammadiha2011b}, this approach suggests a solid
framework to incorporate temporal dynamics into the system. Also, the computational complexity of this method is significantly less than
\cite{Mysore2011}.

Raj \textit{et al.} \cite{Raj2011} proposed a phoneme-dependent approach to use NMF for speech enhancement in which a set of basis vectors are learned
for each phoneme a priori. Given the noisy recording, an iterative NMF-based speech enhancer combined with an automatic speech recognizer
(ASR) is pursued to estimate the clean speech signal. In the experiments, a mixture of speech and music is considered and using a set of speaker-dependent basis matrices the estimation of the clean speech is carried out.

NMF-based noise PSD estimation is addressed in \cite{Mohammadiha2011b}. In this work, the speech and noise basis matrices are trained offline,
after which a constrained NMF is applied to the noisy spectrogram in a frame by frame basis. To utilize the time dependencies of the speech and
noise signals, an $l_{2}$-norm regularization term is added to the cost function. This penalty term encourages consecutive speech and noise NMF coefficients to take similar values, and hence, to model the signals' time dependencies. The instantaneous
noise periodogram is obtained similarly to (\ref{eq:Wiener_filtering}) by switching the role of speech and noise approximates. This estimate
is then smoothed over time using an exponential smoothing to get a less-fluctuating estimate of the noise PSD, which can be combined
with any algorithm that needs a noise PSD, e.g., Wiener filter.

\section{Speech Enhancement Using Bayesian NMF\label{sec:Speech-Enhancement-Using}}
In this section, we present our Bayesian NMF (BNMF) based speech enhancement methods. In the following, an overview of the employed BNMF is provided first, which was originally proposed in \cite{Cemgil2009}. Our proposed extensions of this BNMF to modeling a noisy signal, namely BNMF-HMM and Online-BNMF are given in Subsections \ref{sub:BNMF-HMM-for-Simultanuously} and \ref{sub:Online-Noise-Basis}, respectively. Subsection \ref{sub:Informative-Priors} presents a method to construct informative priors to use temporal dynamics in NMF.

\begin{figure}
\center
\includegraphics[scale=0.5]{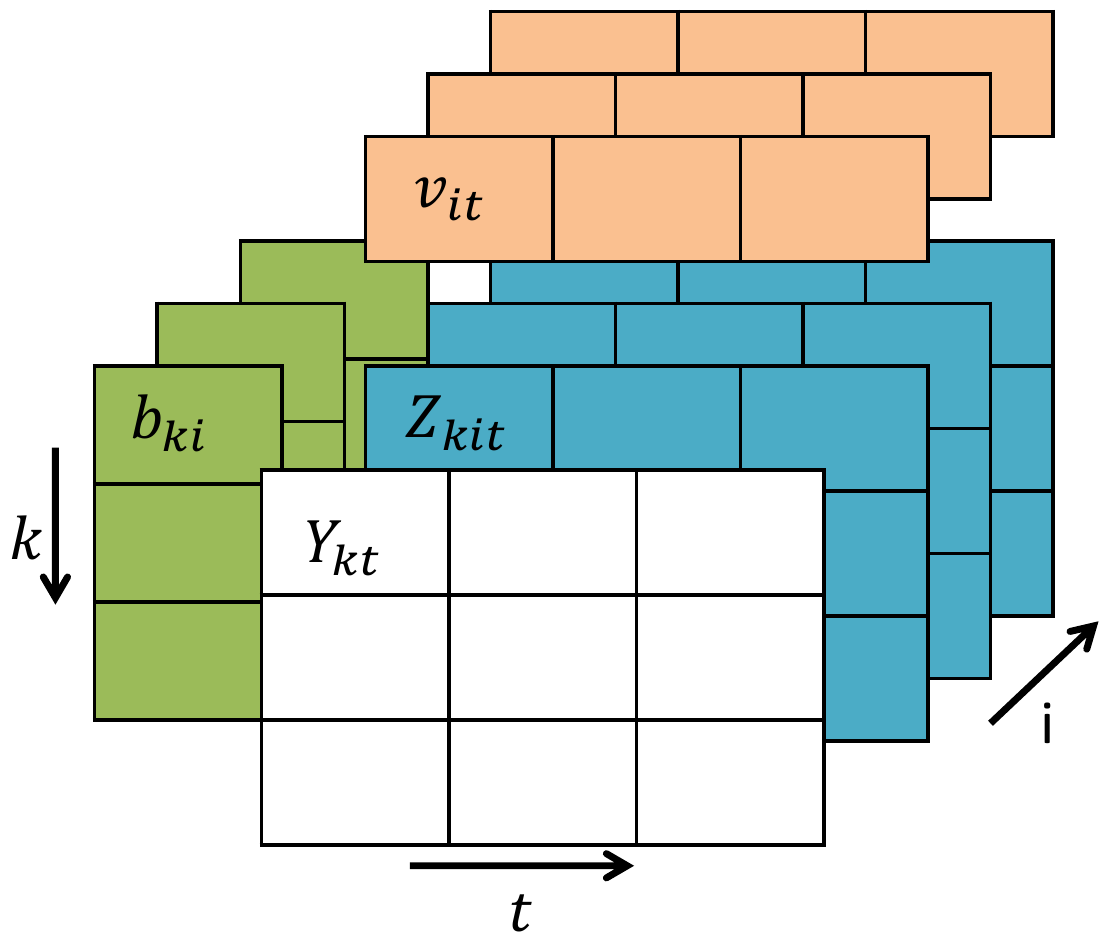}
\caption{\label{fig:bnmf-model} A schematic representation of \eqref{eq:poisson_model1} and \eqref{eq:poisson_model2} \cite{Cemgil2009}. Each time-frequency bin of a magnitude spectrogram ($Y_{kt}$) is assumed to be a sum of some Poisson-distributed hidden random variables ($Z_{kit}$).}
\end{figure}

The probabilistic NMF in \cite{Cemgil2009} assumes that an input matrix is stochastic, and to perform NMF as $\mathbf{y}\approx\mathbf{b}\mathbf{v}$ the following model is considered:
\begin{eqnarray}
Y_{kt} & = & \sum_{i}Z_{kit},\label{eq:poisson_model1}\\
\begin{array}{c}
f_{Z_{kit}}\left(z_{kit}\right)\end{array} & = & \mathit{\mathfrak{\mathcal{PO}}\left(z_{kit};b_{ki}v_{it}\right)}\nonumber \\
 & = & \left(b_{ki}v_{it}\right)^{z_{kit}}e^{-b_{ki}v_{it}}/\left(z_{kit}!\right),\label{eq:poisson_model2}
\end{eqnarray}
where $Z_{kit}$ are latent variables, $\mathcal{PO}\left(z;\lambda\right)$ denotes the Poisson distribution, and $z!$ is the factorial of $z$. A schematic representation of this model is shown in Fig.~\ref{fig:bnmf-model}.

As a result of \eqref{eq:poisson_model1} and \eqref{eq:poisson_model2}, $Y_{kt}$ is assumed Poisson-distributed and integer-valued. In practice, the observed spectrogram is first scaled
up and then rounded to the closest integer numbers to avoid large quantization errors. The maximum likelihood (ML) estimate of the parameters $\mathbf{b}$ and $\mathbf{v}$ can be obtained using an EM algorithm \cite{Cemgil2009}, and the result would be identical to the well-known multiplicative update rules for NMF using Kullback-Leibler (KL-NMF) divergence \cite{Lee2000}.

In the Bayesian formulation, the nonnegative factors are further assumed to be random variables. In this hierarchical model, gamma prior distributions are
considered to govern the basis ($\mathbf{B}$) and NMF coefficient ($\mathbf{V}$) matrices:
\begin{equation}
\begin{array}{c}
f_{V_{it}}\left(v_{it}\right)=\mathfrak{\mathcal{G}}\left(v_{it};\phi_{it},\theta_{it}/\phi_{it}\right),\\
f_{B_{ki}}\left(b_{ki}\right)=\mathfrak{\mathcal{G}}\left(b_{ki};\psi_{ki},\gamma_{ki}/\psi_{ki}\right),
\end{array}\label{eq:prior_structure}
\end{equation}
in which $\mathcal{G}\left(v;\phi,\theta\right)=\exp(\left(\phi-1\right)\log v-v/\theta-\log\Gamma\left(\phi\right)-\phi\log\theta)$ denotes the gamma density function with $\phi$ as the shape parameter and $\theta$ as the scale parameter, and $\Gamma\left(\phi\right)$
is the gamma function. $\phi,\theta,\psi,$ and $\gamma$ are referred to as the hyperparameters.

As the exact Bayesian inference for \eqref{eq:poisson_model1}, \eqref{eq:poisson_model2}, and (\ref{eq:prior_structure}) is difficult, a variational Bayes approach has been proposed in \cite{Cemgil2009} to obtain the approximate posterior distributions
of $\mathbf{B}$ and $\mathbf{V}$. In this approximate inference, it is assumed that the posterior distribution of the parameters are independent, and these uncoupled posteriors are inferred iteratively by maximizing a lower bound on the marginal log-likelihood of data.

More specifically for this Bayesian NMF, in an iterative scheme, the current estimates of the posterior distributions of $\mathbf{Z}$
are used to update the posterior distributions of $\mathbf{B}$ and $\mathbf{V}$, and these new posteriors are used
to update the posteriors of $\mathbf{Z}$ in the next iteration. The iterations are carried on until convergence. The posterior
distributions for $Z_{k,:,t}$ are shown to be multinomial density functions ($:$ denotes 'all the indices'), while for $B_{ki}$ and
$V_{it}$ they are gamma density functions. Full details of the update rules can be found in \cite{Cemgil2009}. This variational approach is much faster than an alternative Gibbs sampler, and its computational complexity can be comparable to that of the ML estimate of the parameters (KL-NMF).

\subsection{BNMF-HMM for Simultaneous Noise Classification and Reduction\label{sub:BNMF-HMM-for-Simultanuously}}
In the following, we describe the proposed BNMF-HMM noise reduction scheme in which the state-dependent output density functions are instances of
the BNMF explained in the introductory part of this section. Each state of the HMM corresponds to one specific noise type. Let us consider a set of noise types for which we are able to
gather some training data, and let us denote the cardinality of the set by $M$. We can train a BNMF model for each of these noise types
given its training data. Moreover, we consider a universal BNMF model for speech that can be trained a priori. Note that the considered speech model doesn't introduce any limitation in the method since we train a model for the speech signal in general, and we don't use any assumption on the identity or gender of the speakers.

\begin{figure}
\center
\includegraphics[scale=0.6]{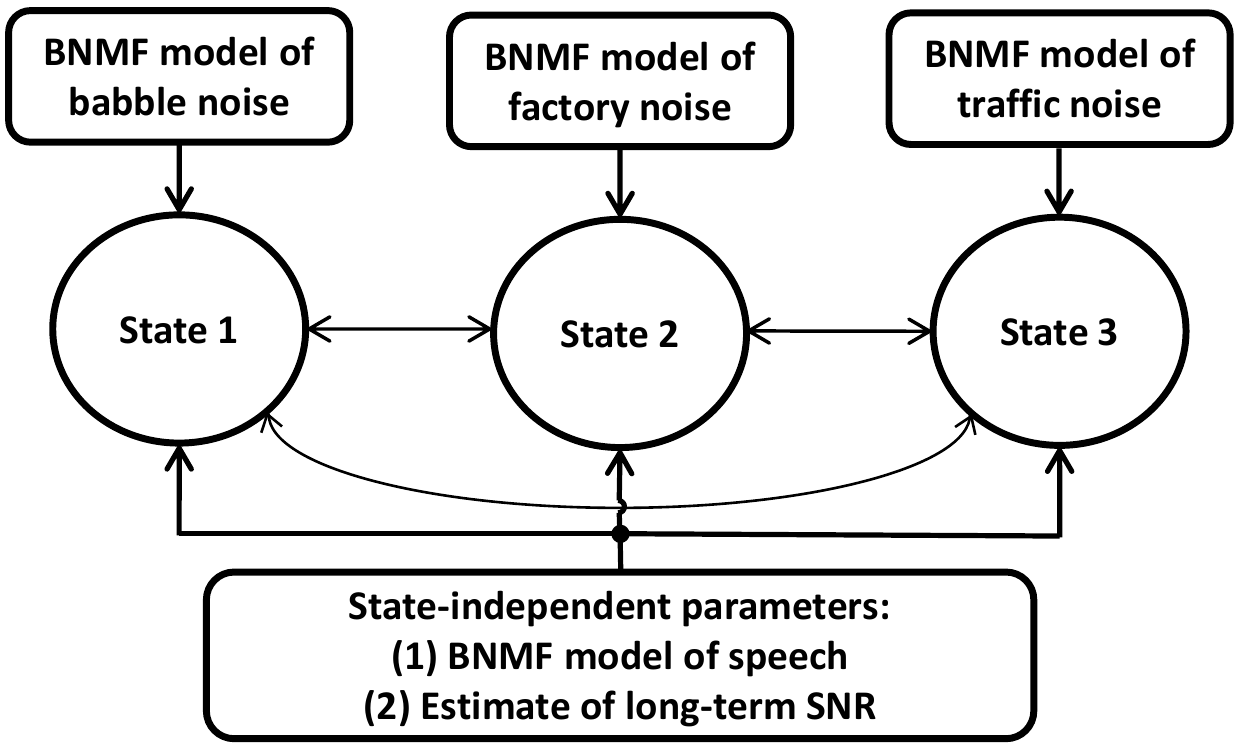}
\caption{\label{fig:BNMF-HMM-structure}A block diagram representation of BNMF-HMM
with three states.}
\end{figure}

The structure of the BNMF-HMM is shown in Fig.~\ref{fig:BNMF-HMM-structure}. Each state of the HMM has some state-dependent parameters, which are
the noise BNMF model parameters. Also, all the states share some state-independent parameters, which consist of the speech BNMF model and an estimate
of the long-term signal to noise ratio (SNR) that will be used for the enhancement. To complete the Markovian model, we need to predefine an empirical state transition matrix (whose dimension is $M\times M$) and an initial state probability vector. For this purpose, we assign some high values to the diagonal elements of the transition matrix, and we set the rest of its elements to some small values such that each row of the transition matrix sums to one. Each element of the initial state probability vector is also set to $1/M$.

We model the magnitude spectrogram of the clean speech and noise signals by \eqref{eq:poisson_model1}. To obtain a BNMF model, we need to find
the posterior distribution of the basis matrix, and optimize for the hyperparameters if desired. During training, we assign some sparse
and broad prior distributions to $\mathbf{B}$ and $\mathbf{V}$ according to (\ref{eq:prior_structure}). For this purpose, $\psi$ and $\gamma$
are chosen such that the mean of the prior distribution for $\mathbf{B}$ is small, and its variance is very high. On the other hand, $\phi$ and $\theta$ are chosen such that the prior distribution of $\mathbf{V}$ has a mean corresponding to the scale of the data and has a high variance
to represent uncertainty. To have good initializations for the posterior means, the multiplicative update rules for KL-NMF are applied first
for a few iterations, and the result is used as the initial values for the posterior means. After the initialization, variational
Bayes (as explained before) is run until convergence. We also optimize the hyperparameters using Newton's method, as proposed in \cite{Cemgil2009}.

In the following, the speech and noise random basis matrices are denoted by ${\mathbf{B}}^{(s)}$ and ${\mathbf{B}}^{(n)}$, respectively. A similar notation is used to distinguish all the speech and noise parameters.

Let us denote the hidden state variable at each time frame $t$ by $X_{t}$, which can take one of the $M$ possible outcomes $x_{t}=1,2,\ldots M$.
The noisy magnitude spectrogram, given the state $X_{t}$, is modeled using (\ref{eq:poisson_model1}). Here, we use the additivity assumption
to approximate the state-dependent distribution of the noisy signal, i.e., $\mathbf{\mathbf{y}}_{t}=\mathbf{s}_{t}+\mathbf{n}_{t}$. To
obtain the distribution of the noisy signal, given the state $X_{t}$, the parameters of the speech and noise basis matrices (${\mathbf{B}}^{(s)}$ and ${\mathbf{B}}^{(n)}$) are concatenated to obtain the parameters of the noisy basis matrix $\mathbf{B}$. Since the sum of independent Poisson random variables is Poisson, (\ref{eq:poisson_model1}) leads to:
\begin{equation}
f_{Y_{kt}}\left(y_{kt}\mid x_t,\mathbf{b},\mathbf{v}_{t}\right)=\frac{\lambda_{kt}^{y_{kt}}e^{-\lambda_{kt}}}{y_{kt}!}\hspace{1mm},
\label{eq:noisy_likelihood}
\end{equation}
where $\lambda_{kt}=\sum_{i}b_{ki}v_{it}$. Note that although the basis matrix $\mathbf{b}$ is state-dependent, to keep the notations uncluttered, we skip writing this dependency explicitly.

The state-conditional likelihood of the noisy signal can now be computed by integrating over $\mathbf{B}$ and $\mathbf{V}_{t}$ as:
\begin{eqnarray}
f_{Y_{kt}}\left(y_{kt}\mid x_{t}\right) & = & \int\int f_{Y_{kt},\mathbf{B},\mathbf{V}_{t}}\left(y_{kt},\mathbf{b},\mathbf{v}_{t}\mid x_{t}\right)d\mathbf{b}\hspace{.2mm}d\mathbf{v}_{t}\nonumber \\
 & = & \int\int f_{Y_{kt}}\left(y_{kt}\mid\mathbf{b},\mathbf{v}_{t},x_{t}\right)\nonumber \\
 &  & f_{\mathbf{B},\mathbf{V}_{t}}\left(\mathbf{b},\mathbf{v}_{t}\mid x_{t}\right)d\mathbf{b}d\mathbf{v}_{t}.\label{eq:joint_distribution}
\end{eqnarray}
The distribution of $\mathbf{y}_{t}$ is obtained by assuming that different frequency bins are independent \cite{Martin2005,Erkelens2007}:
\begin{equation}
f_{\mathbf{Y}_{t}}\left(\mathbf{y}_{t}\mid x_t\right)=\prod_{k}f_{Y_{kt}}\left(y_{kt}\mid x_t\right).
\label{eq:likelihood_total}
\end{equation}

As the first step of the enhancement, variational Bayes approach is applied to approximate the posterior distributions of the NMF coefficient vector $\mathbf{V}_{t}$ by maximizing the variational lower bound on \eqref{eq:likelihood_total}. Here, we assume that the state-dependent posterior distributions of $\mathbf{B}$ are time-invariant and are identical to those obtained during the training. Moreover, we use the temporal dynamics of noise and speech to construct informative prior distributions for $\mathbf{V}_{t}$, which is explained in Subsection \ref{sub:Informative-Priors}.
After convergence of the variational learning, we will have the parameters (including expected values) of the posterior distributions of $\mathbf{V}_{t}$ as well as the latent variables $\mathbf{Z}_{t}$.

The MMSE estimate \cite{Kay1993} of the speech DFT magnitudes can be shown to be \cite{Ephraim1992,Mohammadiha2013c}:
\begin{equation}
\hat{s}_{kt}=E\left(S_{kt}\mid\mathbf{\mathbf{y}}_{t}\right)=\frac{\sum_{x_{t}=1}^{M}\xi_{t}\left(\mathbf{\mathbf{y}}_{t},x_{t}\right)E\left(S_{kt}\mid x_{t},\mathbf{\mathbf{y}}_{t}\right)}{\sum_{x_{t}=1}^{M}\xi_{t}\left(\mathbf{\mathbf{y}}_{t},x_{t}\right)},
\label{eq:speech_estimate}
\end{equation}
where
\begin{eqnarray}
\xi_{t}\left(\mathbf{\mathbf{y}}_{t},x_{t}\right) & = & f_{\mathbf{Y}_{t},X_{t}}\left(\mathbf{\mathbf{y}}_{t},x_{t}\mid\mathbf{\mathbf{y}}_{1}^{t-1}\right)\nonumber \\
 & = & f_{\mathbf{Y}_{t}}\left(\mathbf{\mathbf{y}}_{t}\mid x_{t}\right)f_{X_{t}}\left(x_{t}\mid\mathbf{\mathbf{y}}_{1}^{t-1}\right),\label{eq:exi_definition}
\end{eqnarray}
in which
$\mathbf{\mathbf{y}}_{1}^{t-1}=\{\mathbf{\mathbf{y}}_{1},\ldots\mathbf{\mathbf{y}}_{t-1}\}$. Here, $f_{X_t}(x_{t}\mid\mathbf{\mathbf{y}}_{1}^{t-1})$ is computed using the forward algorithm \cite{Bilmes1997}. Since (\ref{eq:joint_distribution}) can not be evaluated analytically, one can either use numerical methods or use approximations to calculate $f_{Y_{kt}}\left(y_{kt}\mid x_t\right)$.  Instead of expensive stochastic integrations, we approximate (\ref{eq:joint_distribution}) by evaluating the integral at the mean value of the posterior distributions of $\mathbf{B}$ and $\mathbf{V}_{t}$:
\begin{equation}
f_{Y_{kt}}\left(y_{kt}\mid x_t\right)\approx f_{Y_{kt}}\left(y_{kt}\mid\mathbf{b}^{\prime},\mathbf{v}_{t}^{\prime},x_t\right),
\label{eq:joint_likelihood_approx}
\end{equation}
where $\mathbf{b}^{\prime}=E(\mathbf{B}\mid\mathbf{y}_{t},x_{t})$, and $\mathbf{v}_{t}^{\prime}=E(\mathbf{V}_{t}\mid\mathbf{y}_{t},x_{t})$
are the posterior means of the basis matrix and NMF coefficient vector that are obtained using variational Bayes. Other types of point approximations have also been used for gain modeling in the context of HMM-based speech enhancement \cite{Zhao2007,Mohammadiha2013a}.

To finish our derivation, we need to calculate the state-dependent MMSE estimate of the speech DFT magnitudes $E(S_{kt}\mid x_{t},\mathbf{\mathbf{y}}_{t})$. First, let us rewrite (\ref{eq:poisson_model1}) for the noisy signal as:
\begin{eqnarray*}
Y_{kt}=S_{kt}+N_{kt} & = & \sum_{i=1}^{{I}^{\left(s\right)}}{Z}^{\left(s\right)}_{kit}+\sum_{i=1}^{{I}^{\left(n\right)}}{Z}^{\left(n\right)}_{kit}=\sum_{i=1}^{{I}^{\left(s\right)}+{I}^{\left(n\right)}}Z_{kit},
\end{eqnarray*}
where ${I}^{(s)}$ and ${I}^{(n)}$ are the number of speech and noise basis vectors, respectively, given $X_{t}$. Then,
\begin{eqnarray}
E\left(S_{kt}\mid x_{t},\mathbf{\mathbf{y}}_{t}\right) & = & E\left(\sum_{i=1}^{{I}^{\left(s\right)}}{Z}^{\left(s\right)}_{kit}\mid x_{t},\mathbf{\mathbf{y}}_{t}\right)\nonumber \\
 & = & \sum_{i=1}^{{I}^{\left(s\right)}}E\left({Z}^{\left(s\right)}_{kit}\mid x_{t},\mathbf{\mathbf{y}}_{t}\right).
 \label{eq:speech_estimate_middle}
\end{eqnarray}
The posterior expected values of the latent variables in (\ref{eq:speech_estimate_middle}) are obtained during variational Bayes and are given by \cite{Cemgil2009}:
\begin{equation}
E\left(Z_{kit}\mid x_{t},\mathbf{\mathbf{y}}_{t}\right)=\frac{e^{E\left(\log B_{ki}+\log V_{it}\mid x_{t},\mathbf{\mathbf{y}}_{t}\right)}}{\sum_{i=1}^{{I}^{\left(s\right)}+{I}^{\left(n\right)}}e^{E\left(\log B_{ki}+\log V_{it}\mid x_{t},\mathbf{\mathbf{y}}_{t}\right)}}y_{kt}.
\label{eq:expected_value_latent}
\end{equation}
Finally, using (\ref{eq:expected_value_latent}) in (\ref{eq:speech_estimate_middle}), we get
\begin{equation}
E\left(S_{kt}\mid x_{t},\mathbf{\mathbf{y}}_{t}\right)=\frac{\sum_{i=1}^{{I}^{\left(s\right)}}e^{E\left(\log B_{ki}+\log V_{it}\mid x_{t},\mathbf{\mathbf{y}}_{t}\right)}}{\sum_{i=1}^{{I}^{\left(s\right)}+{I}^{\left(n\right)}}e^{E\left(\log B_{ki}+\log V_{it}\mid x_{t},\mathbf{\mathbf{y}}_{t}\right)}}y_{kt}.
\label{eq:mmse_speech_sd}
\end{equation}
As mentioned before, the posterior distributions of $\mathbf{B}$ and $\mathbf{V}$ are gamma density functions and the required expected values to evaluate
(\ref{eq:mmse_speech_sd}) are available in closed form. The time-domain enhanced speech signal is reconstructed using \eqref{eq:speech_estimate} and the noisy phase information.

Eq. (\ref{eq:mmse_speech_sd}) includes Wiener filtering (\ref{eq:Wiener_filtering}) as a special case. When the posterior distributions of the basis and NMF coefficients are very sharp (which happens for large shape parameters in the gamma
distribution), $E(\log V_{it}\mid x_{t},\mathbf{\mathbf{y}}_{t})$ approaches the logarithm of the mean value of the posterior distribution,
$\log(E(V_{it}\mid x_{t},\mathbf{\mathbf{y}}_{t}))$. This can be easily verified by considering that for very large arguments the logarithm provides an accurate approximation to the digamma function. Therefore, for large posterior shape parameters (\ref{eq:mmse_speech_sd}) converges asymptotically to (\ref{eq:Wiener_filtering}). In this case, the mean values of the posterior distributions are used to design the Wiener filter.

We can use $\xi_{t}\left(\mathbf{\mathbf{y}}_{t},x_{t}\right)$ to classify the acoustic noise more explicitly. For this purpose, we
compute the posterior state probability as:
\begin{equation}
f\left(x_{t}\mid\mathbf{\mathbf{y}}_{1}^{t}\right)=\frac{f(\mathbf{\mathbf{y}}_{t},x_{t}\mid\mathbf{\mathbf{y}}_{1}^{t-1})}{\sum_{x_{t}}f(\mathbf{\mathbf{y}}_{t},x_{t}\mid\mathbf{\mathbf{y}}_{1}^{t-1})}.
\label{eq:noise_classifier}
\end{equation}

To reduce fluctuations, it is helpful to smooth \eqref{eq:noise_classifier} over time. Other likelihood-based classification techniques have been used in \cite{Sameti1998,Srinivasan2006} for HMM-based and codebook-driven denoising approaches. In \cite{Srinivasan2006}, a long-term noise PSD is computed using a separate noise PSD tracking algorithm and is used to select one of the available noise models to enhance the noisy signal. Alternatively, in \cite{Sameti1998}, a single noise HMM is selected during periods
of speech pauses and is used to enhance the noisy signal until the next speech pause when a new selection is made. Our proposed classification
in (\ref{eq:noise_classifier}) neither needs an additional noise PSD tracking algorithm, nor requires a voice activity detector.

\subsection{Online Noise Basis Learning for BNMF\label{sub:Online-Noise-Basis}}
We present our scheme to learn the noise basis matrix from the noisy data in this subsection. The online-adapted noise basis is then employed to enhance the noisy signal using the BNMF approach, similarly to \ref{sub:BNMF-HMM-for-Simultanuously} with only one state in the HMM. We continue to use a universal speech model that is learned offline.

To update the noise basis, we store $N_1$ past noisy magnitude DFT frames into a buffer $\underline{\mathbf{n}}\in\mathbb{R}_{+}^{K\times N_1}$, where $K$ is the length of $\mathbf{y}_{t}$. The buffer will be
updated when a new noisy frame arrives. Then, keeping the speech basis unchanged, variational Bayes is applied to $\underline{\mathbf{n}}$
to find the posterior distributions of both the speech and noise NMF coefficients and noise basis matrix.

\begin{figure}
\center
\includegraphics[scale=0.47]{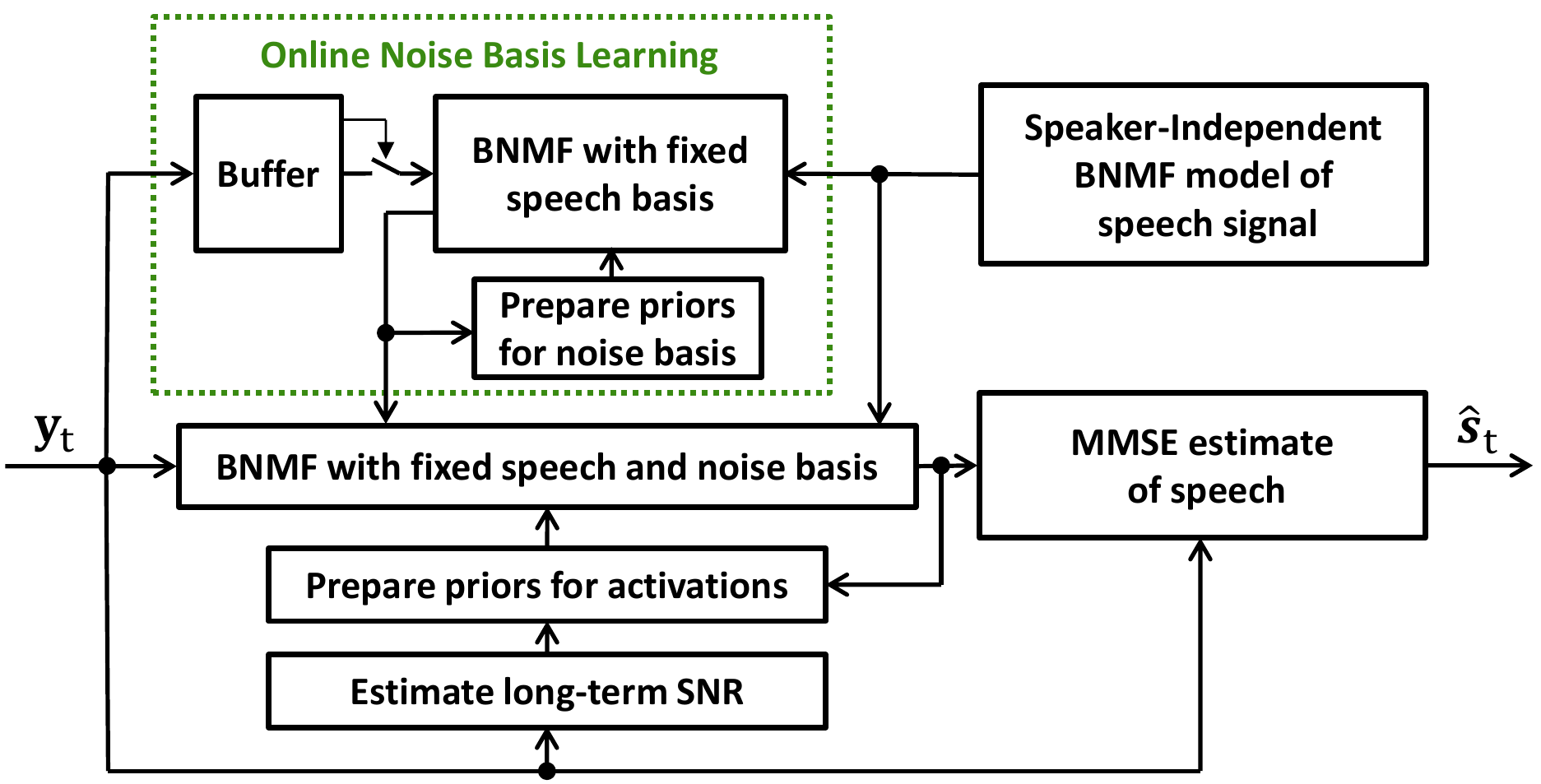}
\caption{\label{fig:Online-BNMF}Block diagram representation of BNMF with online noise basis learning. $\mathbf{y}_{t}$ and $\hat{\mathbf{s}}_{t}$
are the short-time spectral amplitudes of the noisy and enhanced speech signals, receptively, at time frame $t$. The goal of the "Prepare priors" boxes is to recursively update the prior distributions, which will be also discussed in \ref{sub:Informative-Priors}.}
\end{figure}

Let us denote the noise dictionary at time index $t-1$ by $f_{{\mathbf{B}}^{(n)}_{t-1}}({\mathbf{b}}^{(n)}_{t-1}\mid\mathbf{y}_{1}^{t-1})$. To maintain a slowly varying basis matrix, we flatten $f_{{\mathbf{B}}^{(n)}_{t-1}}({\mathbf{b}}^{(n)}_{t-1}\mid\mathbf{y}_{1}^{t-1})$ and use it as the prior distribution for the noise basis matrix at time $t$. Accordingly, using the notation from (\ref{eq:prior_structure}), we set ${\boldsymbol{\gamma}}^{(n)}=E({\mathbf{B}}^{(n)}_{t})=E\left({\mathbf{B}}^{(n)}_{t-1}\mid\mathbf{y}_{1}^{t-1}\right)$,
and ${\psi}^{(n)}_{ki}$ is set to a high value (${\psi}^{(n)}_{ki}={\psi}^{(n)}\gg1,k=1,\ldots K,i=1,\ldots{I}^{(n)})$ to avoid overfitting. With a high value for the shape parameter, the posterior distributions are flattened only slightly to obtain a quite sharp prior distribution. Therefore, the posteriors of the noise basis matrix are encouraged to follow the prior patterns unless the noise spectrogram changes heavily. Fig.~\ref{fig:Online-BNMF} shows a simplified diagram of the online BNMF approach. The top part of the figure (dashed-line box) illustrates the online noise basis learning.

Two points have to be considered to complete the online learning. As we don't expect the noise type to change {rapidly},
we can reduce the computational complexity by updating the noise dictionary less frequently. Also, as an inherent property of NMF, good initializations can improve the dictionary learning. To address these two issues, we use a simple approach based on a sliding window concept. Let us
define a local buffer $\underline{\mathbf{m}}\in\mathbb{R}_{+}^{K\times N_2}$ that stores the last $N_2$ observed noisy DFT magnitudes.
Every time we observe a new frame, the columns in $\underline{\mathbf{m}}$ are shifted to the left and the most recent frame is stored at the
rightmost column. When the local buffer is full, i.e., $N_2$ new frames have been observed, a number of frames (let's say $q$ frames) that
have the lowest energies are chosen to update the main buffer $\underline{\mathbf{n}}$. Note that to do this we don't use any voice activity detector. Hence, the columns in $\underline{\mathbf{n}}$ are shifted to the left and new data is stored on the rightmost columns of the buffer. We now apply the
KL-NMF on $\underline{\mathbf{n}}$ for few iterations, and use the obtained basis matrix to initialize the posterior means of the noise basis
matrix. Then, the iterations of variational Bayes (using both speech and noise basis matrices) are continued until convergence.

{One of the important parameters in our online learning is $N_1$, size of the main buffer. Although a large buffer reduces the overfitting risk, it slows down the adaption speed of the basis matrix. The latter causes the effect of the previous noise to fade out slowly, which will be illustrated in the following example. In our experiments, we set $N_1=50$, $N_2=15$, $q=5$. Our approach of the basis adaption is independent of the underlying SNR.}

\begin{figure}
\center
\includegraphics[scale=0.6]{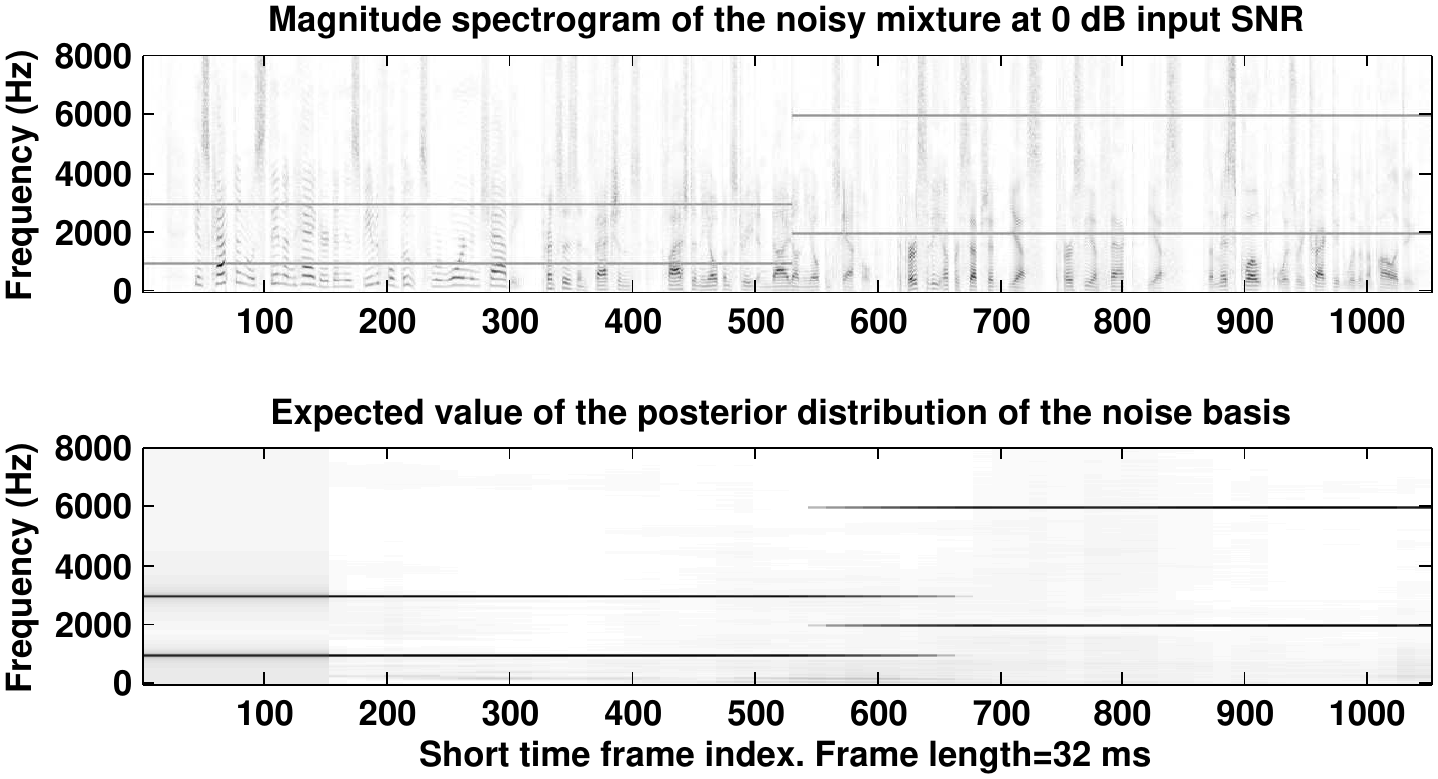}
\caption{\label{fig:Demonstration-onlineBasis}Demonstration of the noise basis
adaptation. The top panel shows a mixture magnitude spectrogram in which a sinusoidal noise signal (having two harmonics corresponding to the horizontal lines) is added to a speech signal at 0 dB input SNR. The bottom panel depicts a single noise basis vector over time that is adapted using the noisy mixture. See the text for more explanation.}
\end{figure}

Fig.~\ref{fig:Demonstration-onlineBasis} provides a demonstration of the online noise basis learning using a toy example. For this example, a noisy signal (at 0 dB SNR) is obtained by adding two different sinusoidal noise signals to the speech waveform at a sampling rate of 16 kHz. A frame length of 32 ms with $50\%$ overlap and a Hann window was utilized to implement the DFT. We learned a single noise basis
vector (${I}^{(n)}=1$) from the noisy mixture. As depicted in the lower panel of Fig.~\ref{fig:Demonstration-onlineBasis}, the noise
basis is adapted correctly to capture the changes in the noise spectrum. BNMF-based speech enhancement resulted to a 13 dB improvement
in source to distortion ratio (SDR) \cite{Vincent2006} and a 0.9 MOS improvement in PESQ \cite{PESQ2000} for this example.

As Fig.~\ref{fig:Demonstration-onlineBasis} demonstrates, the proposed online learning has introduced a latency of around 15 frames in the adaption of the noise basis. In general, this
delay depends on both $N_2$ and the time alignment of the signals, but it is always upper bounded by $2N_2-q$ short-time frames.
Moreover, Fig.~\ref{fig:Demonstration-onlineBasis} shows a side effect of the sliding window where the effect of the previous noise is fed out slowly (depending on the parameters $N_1$, $N_2$ and $q$). However, in a practical scenario, the effect of
this latency and slow decay are not as clear as this toy example because the noise characteristics change gradually and not abruptly.

An additional approach to adapt the noise basis is to update the basis matrix in each short-time frame. In this view, variational Bayes
is applied to each noisy frame to obtain the posterior distribution of both the NMF coefficients and the noise basis matrix. However,
our simulations showed that this approach is not robust enough to changes in the noise type. In fact, to capture the noise spectrogram changes
and at the same time not overfit to a single frame, a tradeoff has to be considered in constructing the priors for the noise dictionary,
which was difficult to achieve in our simulations.

\subsection{Informative Priors for NMF Coefficients\label{sub:Informative-Priors}}
To apply variational Bayes to the noisy signal, we use the temporal dependencies of data to assign prior distributions for the NMF coefficients
$\mathbf{V}$. Both BNMF-based methods from \ref{sub:BNMF-HMM-for-Simultanuously} and \ref{sub:Online-Noise-Basis} use this approach to recursively update the prior distributions. To model temporal dependencies and also to account for the non-stationarity of the signals, we obtain a prior for $\mathbf{V}_{t}$ by widening
the posterior distributions of $\mathbf{V}_{t-1}$. Recalling (\ref{eq:prior_structure}), let the state-conditional prior distributions
be: $f_{V_{it}}(v_{it}\mid x_{t})=\mathfrak{\mathcal{G}}(v_{it};\phi_{it}[x_{t}],\theta_{it}[x_{t}]/\phi_{it}[x_{t}])$ where state dependency is made explicit through the notation $[x_t]$.  For this gamma distribution we have:
\begin{equation}
E\left(V_{it}\mid x_{t}\right)=\theta_{it}\left[x_{t}\right],\quad\frac{\sqrt{\text{var}\left(V_{it}\mid x_{t}\right)}}{E\left(V_{it}\mid x_{t}\right)}=\frac{1}{\sqrt{\phi_{it}\left[x_{t}\right]}},
\label{eq:prior_param}
\end{equation}
where $\text{var}(\cdot)$ represents the variance. We assign the following recursively updated mean value to the prior distribution:
\begin{gather}
\theta_{it}\left[x_{t}\right]=\alpha\hspace{.2mm}\theta_{i,t-1}\left[x_{t}\right]+\left(1-\alpha\right)E\left(V_{i,t-1}\mid\mathbf{y}_{t-1},x_{t}\right),
\label{eq:prior_MMSE}
\end{gather}
where the value of $\alpha$ controls the smoothing level to obtain the prior. Note that due to the recursive updating, $\theta_{it}$
is dependent on all the observed noisy data $\mathbf{y}_{1}^{t-1}$.

In (\ref{eq:prior_param}), different shape parameters are used for
the speech and noise NMF coefficients, but they are constant over time. Thus, $\phi_{it}=\phi_{i,t-1}=\ldots\phi_{i1}$, also $\phi_{it}={\phi}^{(s)}$
for $i=1,\ldots{I}^{(s)},$ and $\phi_{it}={\phi}^{(n)}$ for $i={I}^{(s)}+1,\ldots{I}^{(s)}+{I}^{(n)}$. Moreover, different noise types are allowed to have different shape parameters. In this form of prior, the ratio between the standard deviation and the expected value is the same for all the NMF coefficients for a source signal. The shape parameter $\phi$ represents the uncertainty of the prior which in turn corresponds to the non-stationarity of the signal being
processed. We can learn this parameter in the training stage using the clean speech or noise signals. Hence, at the end of
the training stage, the shape parameters of the posterior distributions of all the NMF coefficients are calculated and their mean value is
taken for this purpose. Using this approach for the speech signal results in ${\phi}^{(s)}=3\sim5$. However, the noise reduction simulations
suggest that having an uninformative prior for speech (a small value for ${\phi}^{(s)}$) leads to a better performance unless the noise signal
is more non-stationary than the speech signal, e.g., keyboard or machine gun noises. Therefore, in our experiments we used a relatively flat
prior for the speech NMF coefficients (${\phi}^{(s)}\ll1$) that gives the speech BNMF model greater flexibility.

\begin{figure}
\center
\includegraphics[scale=0.5]{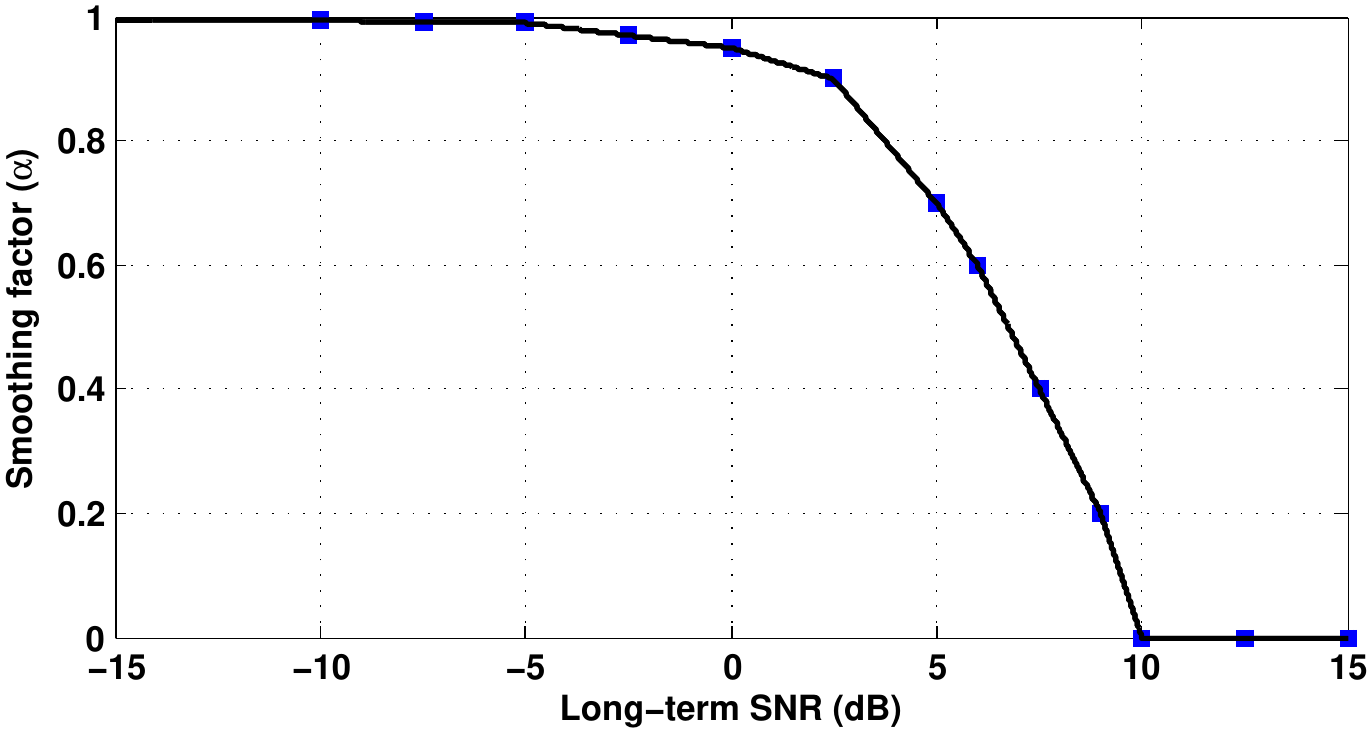}
\caption{\label{fig:SNR-curve}An empirical $\alpha$-SNR curve, which is used in our experiments. The figure shows that for low input SNRs (high noise levels) a high degree of smoothing should be applied to update the mean values of the prior distributions for NMF coefficients \eqref{eq:prior_MMSE}, and vice versa.}
\end{figure}

Our experiments show that the optimal amount of smoothing in (\ref{eq:prior_MMSE}) depends on the long-term SNR (or global SNR). For low SNRs
(high level of noise) a strong smoothing ($\alpha\rightarrow1$) improves the performance by reducing unwanted fluctuations while for high SNRs
a milder smoothing ($\alpha\rightarrow0$) is preferred. The latter case corresponds to obtaining the mean value $\theta$ directly using the information
from the previous time frame. Here, in contrast to \cite{Mohammadiha2012a}, we use an SNR-dependent value for the smoothing factor. Fig.~\ref{fig:SNR-curve} shows an $\alpha-\text{SNR}$ curve that we obtained using computer simulations and was used in our experiments.

{To calculate the long-term SNR from the noisy data, we implemented the approach proposed in \cite{Kim2008} that works well enough for our purpose. This approach assumes that the amplitude of the speech waveform is gamma-distributed with a shape parameters fixed at 0.4, and that the background noise is Gaussian-distributed, and that speech and noise are independent. Under these assumptions, authors have modeled the amplitude of the noisy waveform with a gamma distribution and have shown that the maximum likelihood estimate of the shape parameter is uniquely determined from the long-term SNR \cite{Kim2008}.}

\section{Experiments and Results\label{sec:Experiments-and-Results}}
We evaluate and compare the proposed NMF-based speech enhancement systems in this section. The experiments are categorized as supervised and unsupervised speech enhancement methods. In Subsection \ref{sub:Noise-Reduction-Using-a-Priori}, we evaluate the noise reduction systems where for each noise type we have access to some training data. Evaluation of the unsupervised denoising schemes is presented in \ref{sub:Unsupervised-Noise-Reduction}, where we assume that we don't have training data for some of the noise types.

In our simulations, all the signals were down-sampled to 16 kHz and the DFT was implemented using a frame length of $512$ samples and
$0.5$-overlapped Hann windows.  The core test set of the TIMIT database ($192$ sentences) \cite{Garofolo1993} was exploited for the noise reduction evaluation. The signal synthesis was performed using the overlap-and-add procedure. SNR was

For all the BNMF-based methods, a universal speaker-independent speech model with 60 basis vectors is learned
using the training data from the TIMIT database. The choice of dictionary size is motivated by our previous study \cite{Mohammadiha2011}. {Moreover, for the BNMF-based approaches the long-term SNR was estimated using \cite{Kim2008} and we used Fig.~\ref{fig:SNR-curve} to apply an SNR-dependent smoothing to obtain the priors.}

As reviewed in Section \ref{sec:Review-of-State-of-the-art}, the method introduced in \cite{Mysore2011} factorizes the whole spectrogram
of the noisy signal, and therefore, is not causal. In order to make it more practical, we considered two causal extensions of this work
and evaluated their performance in this section. The first extension is a supervised approach that works frame by frame. Here, we trained one universal NHMM (100 states and 10 basis vectors per state) for speech and one single-state NHMM for each noise type. To achieve causality, we simply replaced the forward-backward algorithm with the forward algorithm in which the NMF coefficients from
the previous timestamp were used to initialize the current ones. As the other extension, we adapted an online noise dictionary learning,  similarly to Section \ref{sub:Online-Noise-Basis}.

\subsection{Noise Reduction Using a-Priori Learned NMF Models\label{sub:Noise-Reduction-Using-a-Priori}}
We evaluated five variants of NMF-based enhancement methods for three noise types. The considered noise types include factory and babble noises from the NOISEX-92 database \cite{noisex92} and city traffic noise from Sound Ideas \cite{Nimens}. Although all of these three noises are considered non-stationary, the city traffic noise is very non-stationary since it includes mainly horn sounds. We implemented five NMF-based algorithms including:
\begin{enumerate}
\item BNMF-HMM: we used (\ref{eq:speech_estimate}) in which the noise-type is not known in advance.

\item General-model BNMF: we trained a single noise dictionary by applying BNMF on a long signal obtained by concatenating the training data of all three noises. For the enhancement, (\ref{eq:mmse_speech_sd}) was used regardless of the underlying noise type.

\item Oracle BNMF: this is similar to BNMF-HMM but the only difference is that instead of the proposed classifier an oracle classifier is used to choose a noise model for enhancement, i.e., the noise type is assumed to be known a priori and its offline-learned basis matrix is used to enhance the noisy signal. Therefore, this approach is an ideal case of BNMF-HMM.

\item Oracle ML: this supervised method is the maximum likelihood implementation of the Oracle BNMF in which KL-NMF in combination with (\ref{eq:Wiener_filtering}) is used to enhance the noisy signal. Similar to the previous case, an oracle classifier is used to choose a noise model for enhancement. The term ML reflects the fact that KL-NMF arises as the maximum likelihood solution of \eqref{eq:poisson_model1} and \eqref{eq:poisson_model2}.

\item Oracle NHMM: this is basically the supervised causal
NHMM, as explained earlier in \ref{sec:Experiments-and-Results}. Similar to cases (3) and (4), the noise type is assumed to be known in advance.
\end{enumerate}
The number of basis vectors in the noise models were set using simulations performed on a small development set. For BNMF and KL-NMF methods, we trained
100 basis vectors for each noise type. Also, 200 basis vectors were learned for the general noise model. For NHMM, a single state with
100 basis vectors were learned for factory and city traffic noises while 30 basis vectors were pre-trained for babble noise since it
provided a better performance.

The performance of the NMF-based methods is compared to a speech short-time spectral amplitude estimator using super-Gaussian prior distributions
\cite{Erkelens2007}, which is referred to as STSA-GenGamma. Here, we used \cite{Hendriks2010} to track the noise PSD, and we set $\gamma=\nu=1$
since it is shown to be one of the best alternatives \cite{Erkelens2007}. This algorithm is considered in our simulations as a state-of-the-art
benchmark to compare NMF-based systems.

\begin{figure}
\center
\includegraphics[scale=0.5]{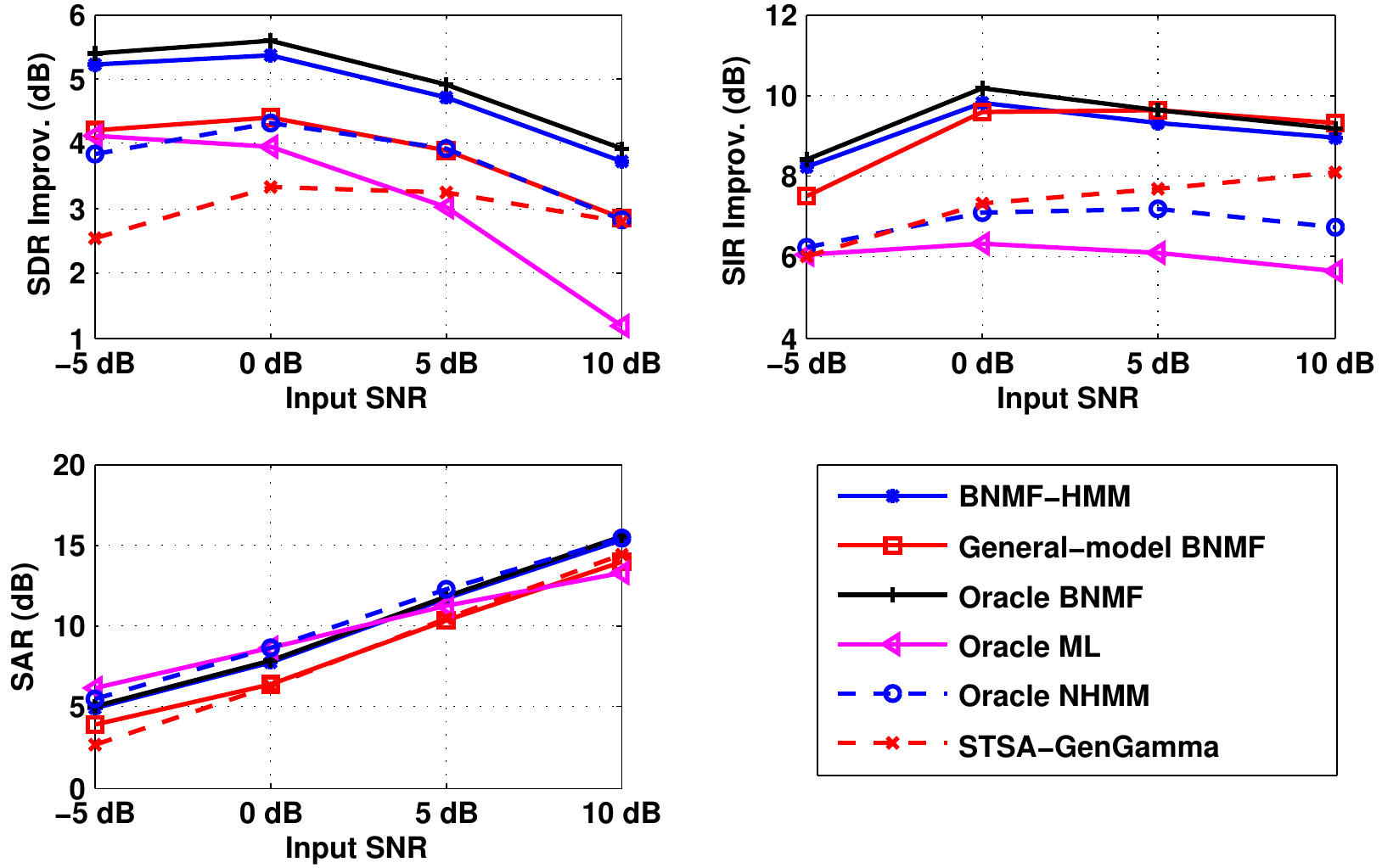}
\caption{\label{fig:BSS-EVAl supervised}BSS-Eval measures \cite{Vincent2006} to evaluate and compare the supervised NMF-based denoising algorithms.  The BNMF-based schemes are described in Subsection \ref{sub:BNMF-HMM-for-Simultanuously}. Here, the prefix "Oracle" is used for the variants where the noise type is known a priori. The results are averaged over different noise types. For the SDR and SIR, improvements gained by the enhancement systems are shown.}
\end{figure}

Fig.~\ref{fig:BSS-EVAl supervised} shows the source to distortion ratio (SDR), source to interference ratio (SIR), and source to artifact
ratio (SAR) from the BSS-Eval toolbox \cite{Vincent2006}. SDR measures the overall quality of the enhanced speech while SIR and SAR are proportional
to the amount of noise reduction and inverse of the speech distortion, accordingly. For SDR and SIR, the improvements gained by the noise reduction systems
are shown. Several interesting conclusions can be drawn from this figure.

The simulations show that the Oracle BNMF has led to the best performance, which is closely followed by BNMF-HMM. The performance of these two
systems is quite close with respect to all three measures. This shows the superiority of the BNMF approach, and also, it indicates that
the HMM-based classification scheme is working successfully. Another interesting result is that except for the Oracle ML, the other NMF-based
techniques outperform STSA-GenGamma. The ML-NMF approach gives a poor noise reduction particularly at high input SNRs. {These results were confirmed by our informal listening tests.}

Moreover, the figure shows that the Oracle NHMM and General-model BNMF methods lead to similar SDR values. However, these two methods
process the noisy signal differently. The NHMM method doesn't suppress a lot of noise but it doesn't distort the speech signal either (i.e.,
SAR is high). This is reversed for the General-model BNMF. Furthermore, comparing BNMF-HMM and General-model BNMF confirms an already reported observation \cite{Sameti1998,Srinivasan2006} that using many small noise-dependent models is superior to a large noise-independent model.

\begin{figure}
\center
\includegraphics[scale=0.5]{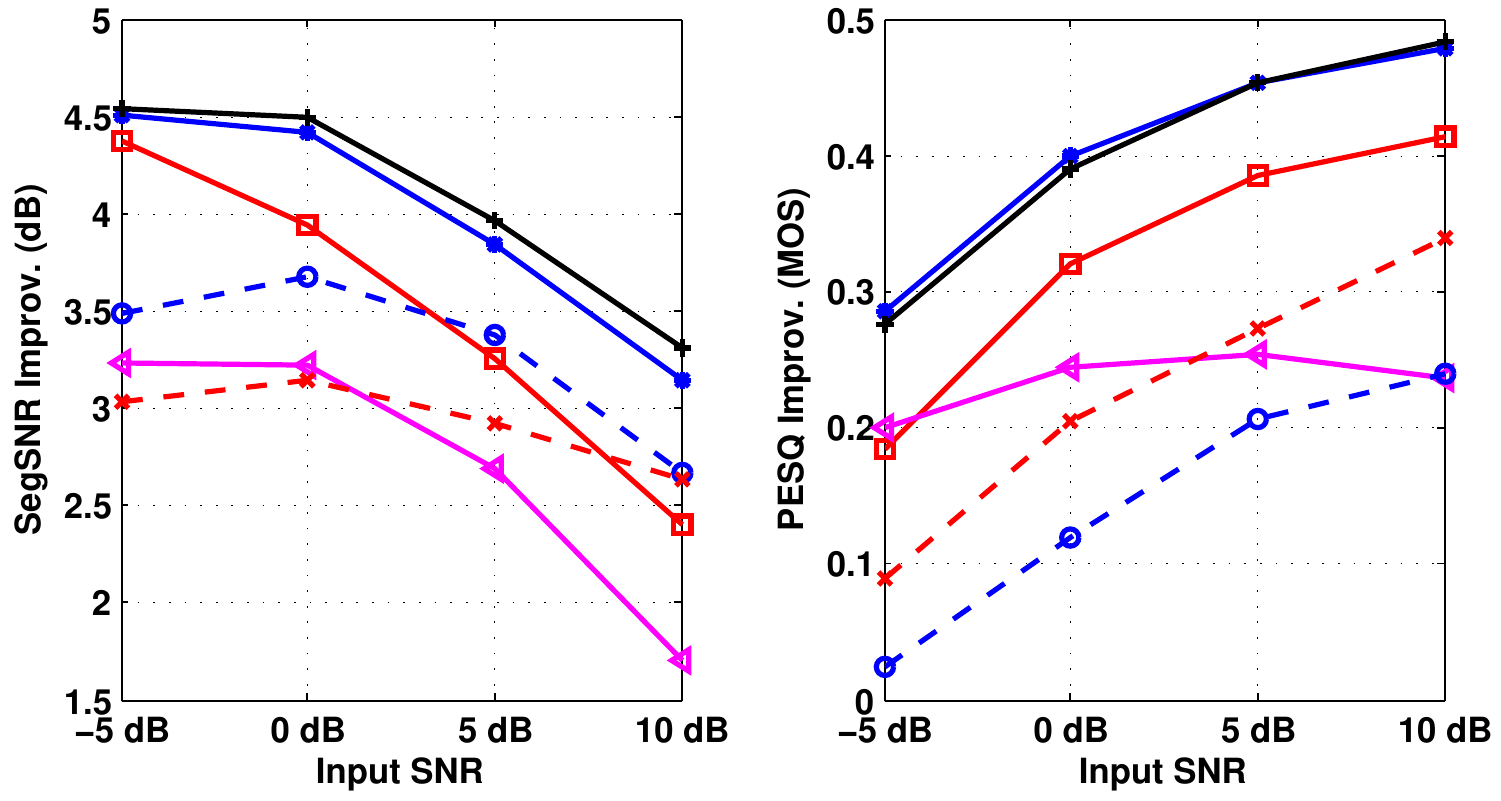}
\caption{\label{fig:pesq-supervised}PESQ and Segmental SNR (SegSNR) improvements gained by the supervised enhancement systems. Legend of this figure is similar
to that of Fig.~\ref{fig:BSS-EVAl supervised}.}
\end{figure}

Fig.~\ref{fig:pesq-supervised} provides the experimental results using segmental SNR (SegSNR) \cite[ch. 10]{Loizou2007}, which is
limited to the range $[-10\text{dB},30\text{dB}]$, and perceptual evaluation of speech quality (PESQ) \cite{PESQ2000}. As it can be
seen in the figure, the BNMF-based methods have led to the highest SegSNR and PESQ improvements. These results verify again the excellence
of the BNMF strategies. Moreover, it is interesting to note that the NHMM method has not been very successful in improving the quality of the noisy
speech with respect to the PESQ measure.

\begin{figure}
\center
\includegraphics[scale=0.58]{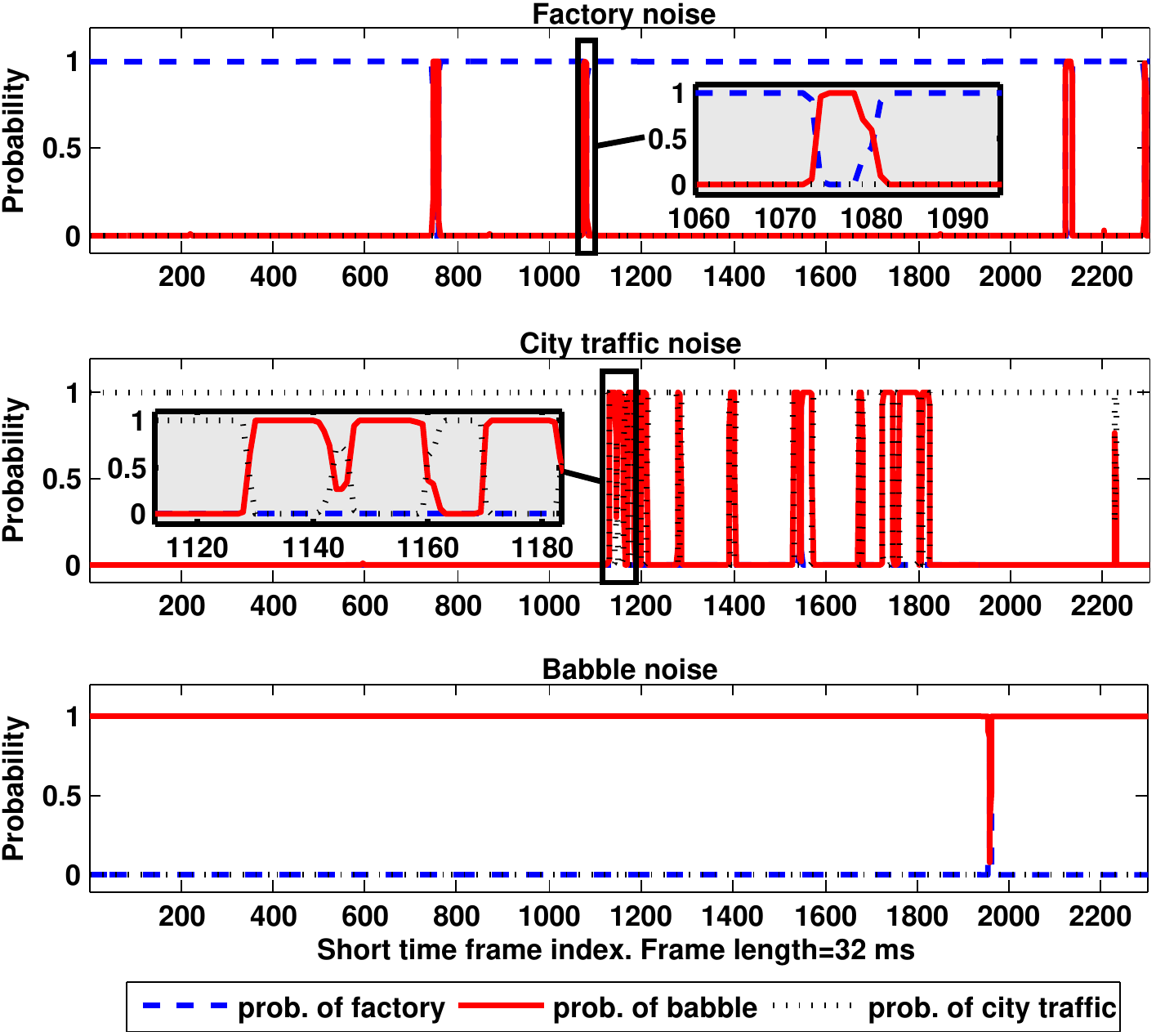}
\caption{\label{fig:classifier} Result of the noise classifier where \eqref{eq:noise_classifier} is smoothed over time and is plotted for a mixture at 0 dB input SNR. The underlying noise type is given in the titles of the subplots (which corresponds to factory, city traffic, and babble noises, respectively, from top to bottom). In each subplot, the probability of three noise classes (factory, city traffic, and babble noises) are shown. For visibility, two small segments are magnified and shown in the figure.}
\end{figure}

To study specifically the classification part of the BNMF-HMM algorithm, we analyzed the output of the classifier. Fig.~\ref{fig:classifier}
provides the result of this experiment. To have a clearer representation, the probability of each noise type in (\ref{eq:noise_classifier}) is smoothed over time and is depicted in Fig.~\ref{fig:classifier}. Here, the classifier is applied to a noisy signal at 0 dB input SNR. The underlying noise type is given as the titles of the subplots. As it can be seen in the figure, the classifier works reasonably well in general. Most of the wrong classifications correspond to the case
where the true noise type is confused with the babble noise. One reason for this confusion is due to the nature of babble noise. If the
short-time spectral properties of the noise are not very different from those of babble, the union of speech and babble basis vectors
can explain any noisy signal by providing a very good fit to the speech part. However, as shown in Fig.~\ref{fig:BSS-EVAl supervised} and
Fig.~\ref{fig:pesq-supervised}, this confusion has reduced the performance only very marginally.

\subsection{Experiments with Unsupervised Noise Reduction\label{sub:Unsupervised-Noise-Reduction} }
This subsection is devoted to investigating the performance of the unsupervised NMF-based speech enhancement systems. For this purpose, we considered
6 different noise types including factory and babble noises from the NOISEX-92 database \cite{noisex92}, and city traffic, highway traffic, ocean,
and hammer noises from Sound Ideas \cite{Nimens}. Among these, ocean noise can be seen as a stationary signal in which the noise level
changes up to $\pm20$ dB. All the signals were concatenated before processing.

We evaluated three NMF-based enhancement systems using a general speech model, which is learned similarly to Subsection \ref{sub:Noise-Reduction-Using-a-Priori}. We considered Online BNMF (Subsection \ref{sub:Online-Noise-Basis})
and Online NHMM (as explained earlier in Section \ref{sec:Experiments-and-Results}). Additionally,
we included the BNMF-HMM in the comparison. The considered BNMF-HMM model was identical to that of Subsection \ref{sub:Noise-Reduction-Using-a-Priori}, i.e., we learned only three models for factory, babble and city traffic noises. For the other noise types, the method is allowed to use any
of these models to enhance the noisy signal according to (\ref{eq:speech_estimate}). Furthermore, we included two state-of-the-art approaches in our experiments: The STSA-GenGamma approach, identical to that of Subsection \ref{sub:Noise-Reduction-Using-a-Priori},
and a Wiener filter in which the noise PSD was estimated using \cite{Hendriks2010} and a decision-directed approach \cite{Ephraim2005} was used to implement the filter. Here, the final gain applied to the noisy signal was limited to be larger than 0.1, for perceptual reasons \cite{Malah1999}.

{For the online BNMF and online NHMM algorithms, we learned ${I}^{(n)}=30$ basis vectors for noise. Learning a large basis matrix is this case can lead to overfitting since the dictionary is adapted given a small number of observations ($N_1=50$ in our experiments). This was also verified in our computer simulations. Hence, in contrast to the supervised methods for which we learned 100 basis vectors for each noise, we learned a smaller dictionary for online algorithms.}

\begin{figure}
\center
\includegraphics[scale=0.5]{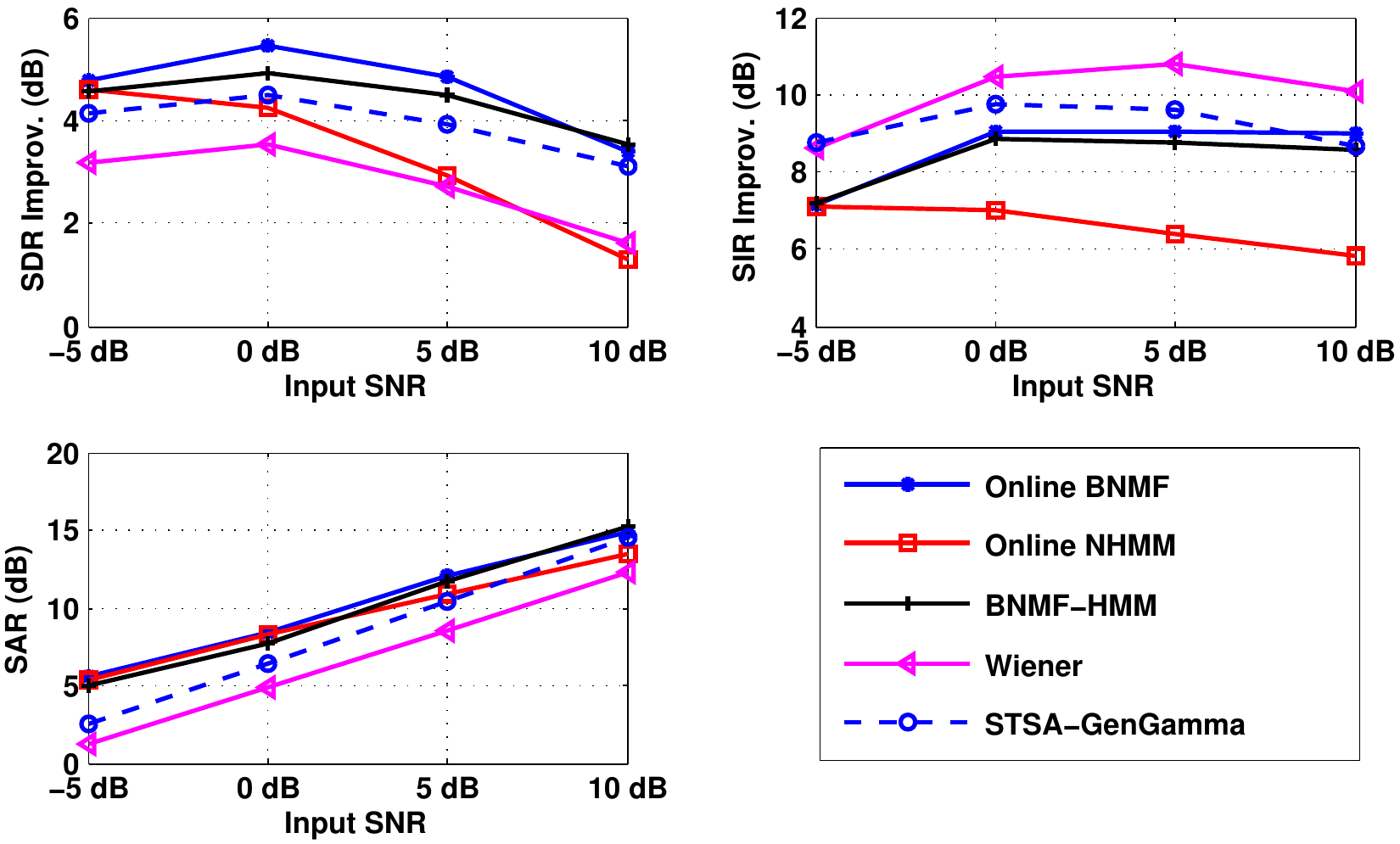}
\caption{\label{fig:bss_unsupervised}{SDR and SIR improvements and SAR measure} \cite{Vincent2006} to evaluate and compare the unsupervised NMF-based denoising algorithms. For the Online BNMF and Online NHMM variants, the noise basis matrix is learned online from the noisy data, explained in \ref{sub:Online-Noise-Basis}. The results are averaged over different noise types. For the BNMF-HMM approach, similar to Fig. \ref{fig:BSS-EVAl supervised}, only three noise models are learned.}
\end{figure}

Fig.~\ref{fig:bss_unsupervised} shows the objective measures from BSS-Eval \cite{Vincent2006} for different algorithms. As it can be
seen in the figure, Online BNMF has outperformed all the other systems. This method introduces the least distortion in the enhanced speech
signal while performing moderate noise reduction. On the other hand, Wiener filter and STSA-GenGamma reduce the interfering noise greatly with the cost
of introducing artifacts in the output signal.

Online NHMM outperforms the Wiener and STSA-GenGamma algorithms at low input SNRs with respect to SDR but for high input SNRs the performance
of the algorithm is the worst among all the competing methods. Also, the amount of noise suppression using Online NHMM is the least among different methods.

Moreover, Fig.~\ref{fig:bss_unsupervised} shows that STSA-GenGamma provides a higher-quality enhanced speech signal than the Wiener filter.
This is reported frequently in the literature, e.g. \cite{Erkelens2007}.

\begin{figure}
\center
\includegraphics[scale=0.5]{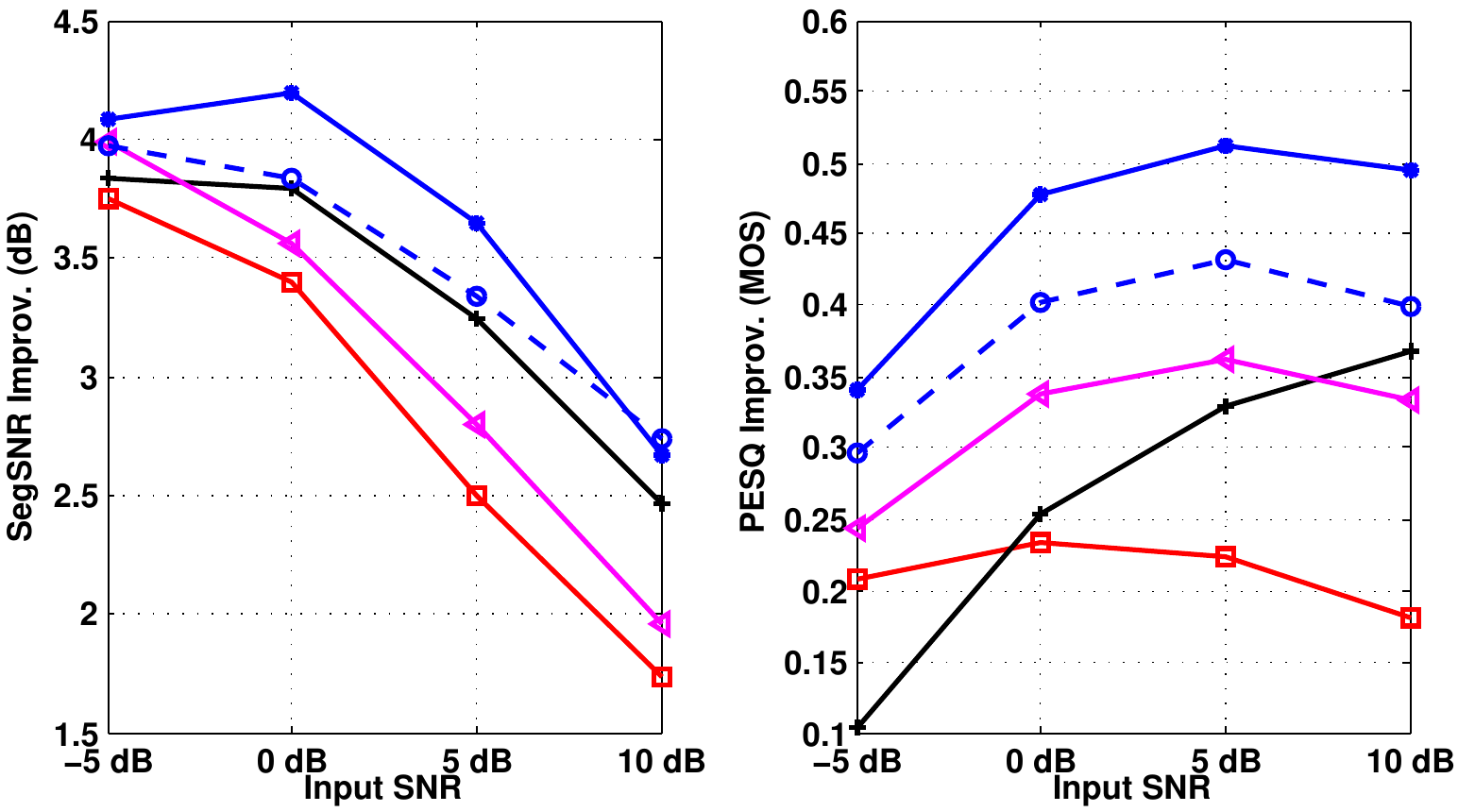}
\caption{\label{fig:PESQ-unsupervised}PESQ and Segmental SNR (SegSNR) improvements gained by the unsupervised enhancement systems. Legend of this figure is similar
to that of Fig.~\ref{fig:bss_unsupervised}.}
\end{figure}

Another interesting result that can be seen in Fig.~\ref{fig:bss_unsupervised} is that Online BNMF outperforms the BNMF-HMM. The difference in the
performance is even larger with respect to SegSNR and PESQ, shown in Fig.~\ref{fig:PESQ-unsupervised}. As it is shown in this
figure, Online BNMF outperforms the BNMF-HMM (and the other methods) with a large margin.

\begin{figure}
\center
\includegraphics[scale=0.55]{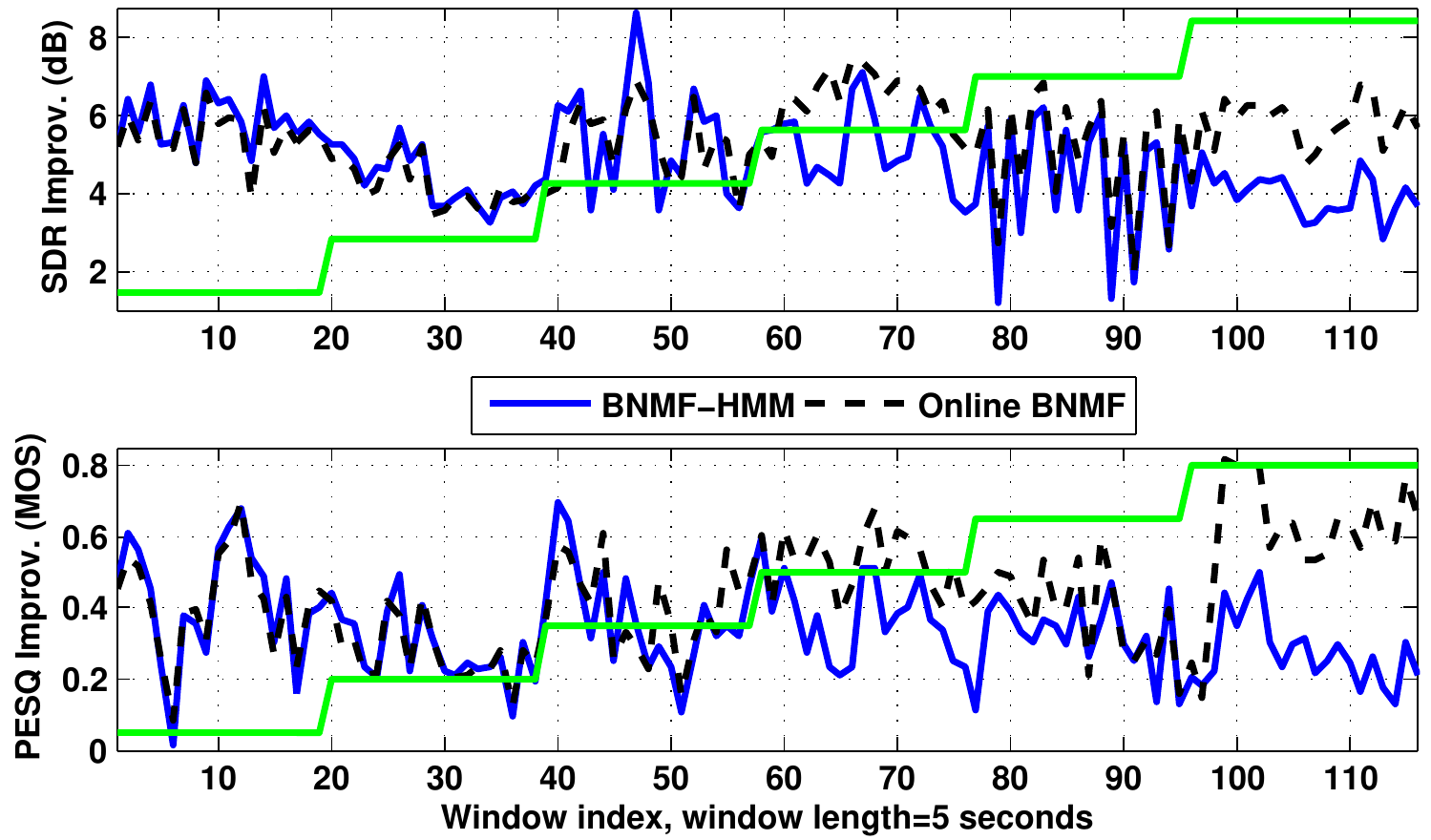}
\caption{\label{fig:windowed_results}SDR and PESQ measured over short intervals of 5-second long. Six different levels shown in green correspond to factory, babble, city traffic, highway traffic, ocean, and hammer noises, respectively
from left to right. For the BNMF-HMM approach, only three noise models corresponding to the first three noises are learned; for the other noise types, the estimator chooses a model that can describe the noisy observation better than the other models. }
\end{figure}

To have a better understanding on how Online BNMF and BNMF-HMM schemes behave for different noise types, we evaluated SDR and PESQ over short intervals of time.
To do so, the noisy and enhanced speech signals were windowed into segments of 5 seconds and then for each segment a SDR and PESQ value was calculated. Fig.~\ref{fig:windowed_results} shows such results as a function of window index. The boundary of the underlying noise types
is shown in green in six different levels in which segments belong to factory, babble, city traffic, highway traffic, ocean, and hammer noises, respectively from
left to right. As can be seen in the figure, for the first three noise types for which a noise-dependent BNMF model is learned offline the
BNMF-HMM approach works marginally better than the Online BNMF. But, for the last three noise types Online BNMF outperforms BNMF-HMM significantly.
The difference is highest for the hammer noise;  this is due to our observation that the hammer noise differs more from either factory, babble or city traffic noises than highway traffic or ocean noises do. Therefore, neither of the pre-trained models can explain the hammer noise well, and as
a result, the overall performance of the BNMF-HMM degrades whenever there is a large mismatch between the training and the
testing signals.

A final remark about the Online BNMF and BNMF-HMM can be made considering the computational complexity. In our simulations (where we didn't use
parallel processing techniques), Online BNMF runs twice as fast as BNMF-HMM with three states. Moreover, our Matlab implementation of
the Online BNMF runs in approximately 5-times real time in a PC with 3.8 GHz Intel CPU and 2 GB RAM.
\section{Conclusions\label{sec:Conclusions}}
This paper investigated the application of NMF in speech enhancement systems. We developed speech enhancement methods using a Bayesian formulation of NMF (BNMF). We proposed two BNMF-based systems to enhance the noisy signal in which the noise type is not known a priori. We developed an HMM in which the output distributions are assumed to be BNMF (BNMF-HMM). The developed method performs a simultaneous noise classification and speech enhancement and therefore doesn't require the noise type in advance. Another unsupervised system was constructed by learning the noise BNMF model online, and is referred to as Online BNMF.

Our experiments showed that a noise reduction system using a maximum likelihood (ML) version of NMF---with a universal speaker-independent speech model----doesn't outperform state-of-the-art approaches. However, by incorporating the temporal dependencies in form of prior distributions and using optimal MMSE filters, the performance of the NMF-based methods increased considerably. The Online BNMF method is faster than the BNMF-HMM and was shown to be superior when the
underlying noise type was not included in the training data. Our simulations showed that the suggested systems outperform the Wiener filter and an
MMSE estimator of speech short-time spectral amplitude (STSA) using super-Gaussian priors with a high margin while they are not restricted to know any priori
information that is difficult to obtain in practice.
\section*{Acknowledgment}
The authors are grateful to Gautham J. Mysore for providing a Matlab implementation of the NHMM approach in \cite{Mysore2011}.
\ifCLASSOPTIONcaptionsoff
  \newpage
\fi
\bibliographystyle{IEEEtran}
{
\bibliography{NasserRefs}


\end{document}